\documentclass[a4paper,11pt]{article}
 \pdfoutput=1

 \usepackage{amsmath, amssymb, latexsym, amscd, amsthm,amsfonts,amstext}
 \usepackage[mathscr]{eucal}
 \usepackage{graphicx}
 \usepackage{subfig}

 \newtheorem{theorem}{Theorem}[section]

    \newtheoremstyle{example}{\topsep}{\topsep}%
      {}
      {}
      {\bfseries}
      {.}
      {5pt}
      {\thmname{#1}\thmnumber{ #2}\thmnote{ #3}}

    \theoremstyle{example}

\def\bR{\mathbb{R}}

\title{Reconstruction of the refractive index from experimental
backscattering data using a globally convergent inverse method}

\author{Nguyen Trung Th\`anh$^a$, Larisa Beilina$^b$, Michael V.~Klibanov$^a$ and Michael A.~Fiddy$^c$\\
\\
$^a$Department of Mathematics \& Statistics, University of North Carolina at Charlotte,\\
Charlotte 28223, NC, USA (tnguy152@uncc.edu; mklibanv@uncc.edu).\\
$^b$Department of Mathematical Sciences, \\
Chalmers University of Technology and
Gothenburg University,\\
 SE-42196 Gothenburg, Sweden (larisa@chalmers.se).\\
$^c$Optoelectronics Center, Univeristy of North Carolina at Charlotte, \\
Charlotte NC 28223, USA. (mafiddy@uncc.edu)}
\date{}

\begin{document}

 \maketitle

\begin{abstract}
The problem to be studied in this work is within the context of coefficient
identification problems for the wave equation. More precisely, we consider
the problem of reconstruction of the refractive index (or equivalently, the
dielectric constant) of an inhomogeneous medium using one backscattering
boundary measurement. The goal of this paper is to analyze the performance
of a globally convergent algorithm of Beilina and Klibanov on experimental
data acquired in the Microwave Laboratory at University of North Carolina at
Charlotte. The main challenge working with experimental data is the the huge
misfit between these data and computationally simulated data. We present
data pre-processing steps to make the former somehow look similar to the
latter. Results of both non-blind and \textit{\textbf{blind}} targets are
shown indicating good reconstructions even for high contrasts between the
targets and the background medium.
\end{abstract}


\textbf{AMS classification codes:} 35R30, 35L05, 78A46.

\section{Introduction}

\label{sec:int}

In this paper, we consider the problem of the reconstruction of the
refractive indices (equivalently, the relative permittivities, or dielectric
constants) of unknown targets placed in a homogeneous medium using
experimental measurements of back-scattered electromagnetic waves in time
domain. Mathematical speaking, this is a coefficient inverse problem (CIP)
for the time-dependent wave-like equation: we reconstruct a spatially
varying coefficient of this equation using measurements on a part of the
boundary of the domain of interest. Potential applications of this problem
are in the detection and characterization of explosives, including
improvised explosive devices (IEDs). Note that IEDs are often located above
the ground surface \cite{K-B-K-S-N-F:IP2012}, which is somewhat close to our
case of targets located in the air. The case when targets are buried in the
ground will be reported in a future publication.

Different imaging methods have been applied to this type of measurements to
obtain geometrical information such as shapes, sizes and locations of the
targets, see e.g.~\cite{Isakov:IP2009,Soumekh:1999}. However, the refractive indices,
which characterize the targets in terms of their constituent materials, are
much more difficult to estimate. This is a motivation of the current
publication.

For conventional gradient-based optimization approaches, there is a huge
literature, see e.g.~\cite{BK:2004,Chavent:2009,E-H-N:1996} and
the references therein. It is well-known that their convergence is
guaranteed only if the starting point of iterations is chosen to be
sufficiently close to the correct solution. This means that they require
some \textit{a priori} information about the targets being found, which is
not always available in many practical situations. Unlike these, a different
method was proposed in \cite{B-K:IP2010, B-K:JIIP2012, K-F-B-P-S:IP2010,
K-B-K-S-N-F:IP2012} and results were summarized in the book \cite{B-K:2012}.
This method provides a good approximation for the exact coefficient without 
\textit{a priori} knowledge of a small neighborhood of this coefficient. Its
global convergence has been rigorously proved for an approximate
mathematical model, see Theorem 2.9.4 in \cite{B-K:2012} and Theorem 5.1 in 
\cite{B-K:JIIP2012}. Due to this model, it is referred to as an \textit{%
approximately globally convergent method} (globally convergent method in
short). In \cite{K-F-B-P-S:IP2010} the authors demonstrated good
reconstruction results for a transmitted experimental data set using this
method, whereas a gradient based method with Tikhonov regularization,
starting from the homogeneous medium as the first guess, failed.

The goal of this paper is to show how this globally convergent method
performs on a backscattering experimental data. While previously it was
demonstrated how this method works on transmitted real data \cite%
{B-K:2012,B-K:IP2010,K-F-B-P-S:IP2010}, the case of backscattering data is
much more complicated, because backscattered signals are much weaker than
transmitted ones.\ In addition, a number of unwanted scattering signals
caused by objects present in the room of experiments\ (e.g. furniture) occur
in the backscattering case. Although in \cite{K-B-K-S-N-F:IP2012}
backscattering data were treated, they were 1-d data only, while we work
here with the 3-d data. To collect these data, an experimental apparatus was
built in the Microwave Laboratory of University of North Carolina at
Charlotte, using support of US Army Research Office.

The main challenge working with our experimental data is a \emph{huge misfit}
between these data and computationally simulated ones, also see \cite%
{B-K:2012,B-K:IP2010,K-F-B-P-S:IP2010,K-B-K-S-N-F:IP2012} for the same
conclusion. From the Functional Analysis standpoint this means that the
function expressing experimental data is far away in any reasonable norm
from the range of the operator of our forward problem. And this operator
should be inverted to solve the inverse problem. Hence, any inversion
algorithm would fail to produce satisfactory results, if being applied to
the raw data.

Therefore, the \emph{central procedure} before applying the globally
convergent algorithm is a heuristic data pre-processing procedure. This
procedure makes the data look somewhat similar to the data provided by
computational simulations. In other words, it moves the data closer to the
range of that operator. The pre-processing of the current paper is
substantially different from pre-processing procedures of \cite%
{B-K:2012,B-K:IP2010,K-F-B-P-S:IP2010,K-B-K-S-N-F:IP2012} because the data
are different. We describe our data pre-processing procedure in section \ref%
{sec:exp}.

The pre-processed data are used as the input for the globally convergent
method. Our goal is to image refractive indices and locations of the
targets. In addition, we want to estimate sizes of those targets. We should
mention that results of this method can be used as
initial guesses for locally convergent methods in order to refine images,
especially the targets' shapes. For example, it was shown in \cite{B-K:2012,B-K:IP2010,BK:AA2013}
that the adaptive finite element method has significantly improved images of
shapes for transmitted experimental data in the case when its starting point
was the solution obtained by the globally convergent method.

The experimental data sets of this paper include both non-blind and blind
cases. \textquotedblleft Blind" means that the targets were unknown for the
computational team (NTT, LB, MVK) but known to MAF, who was leading the data
collection process. Moreover, refractive indices of these targets were
measured after the reconstruction results were obtained. Then computational
results were compared with directly measured ones. Our results indicate that
we not only reconstruct accurately refractive indices and locations of
targets, but also can differentiate between metallic and non-metallic
targets.


\section{Problem statement and the globally convergent method}

\label{sec:model}

In this section we state the forward and inverse problems under
consideration as well briefly outline the globally convergent method of \cite%
{B-K:2012} for reader's convenience. We refer to \cite{B-K:2012} and \cite%
{B-K:JIIP2012} for more details.

\subsection{Forward and inverse problems}

As the forward problem, we consider the propagation of the electromagnetic
wave generated by a point source in $\bR^{3}$. Below, $%
\mathbf{x}=(x,y,z)$ denotes a point in $\bR^{3}$. Since in
our experiments only one component of the electric wave field $E$ is
generated from the transmitting horn antenna (source) and the detector
measures only that component of the scattered electric field, we model the
wave propagation by the following Cauchy problem for the scalar wave
equation 
\begin{eqnarray}
&&\epsilon (\mathbf{x})u_{tt}(\mathbf{x},t)=\Delta u(\mathbf{x},t),(\mathbf{x%
},t)\in \bR^{3}\times (0,\infty ),  \label{eq:fp1} \\
&&u(\mathbf{x},0)=0,\quad u_{t}(\mathbf{x},0)=\delta (\mathbf{x}-\mathbf{x}%
_{0}),  \label{eq:fp2}
\end{eqnarray}%
where $u$ is the total wave which is equal to the sum of the incident wave $%
u^{i}$ and the scattered wave $u^{s}$ caused by the scattering from the
inhomogeneous medium. To further justify our use of the single equation (\ref%
{eq:fp1}) instead of the full Maxwell' system, we refer to \cite%
{Beilina:CEJM2013}, where it was shown numerically that the component of $E$
which was initially incident upon the medium, dominates two other
components, at least for a rather simple medium, such as we work with, see
section 7.2.2 of \cite{Beilina:CEJM2013}. Besides, equation (\ref{eq:fp1})
was previously successfully used to work with transmitted experimental data 
\cite{B-K:2012,B-K:IP2010,BK:AA2013}. In (\ref{eq:fp1}), $\epsilon (\mathbf{x%
})$ represents the spatially distributed dielectric constant, while $n(%
\mathbf{x}):=\sqrt{\epsilon (\mathbf{x})}$ is referred to as the refractive
index of the medium in which the wave propagates. In our analysis, we assume
that the coefficient $\epsilon (\mathbf{x})$ is unknown inside of a bounded
convex domain $\Omega \subset \bR^{3}$ with $\partial \Omega
\in C^{3}.$ We also assume the existence of a positive constant $d$ such
that 
\begin{equation}
\epsilon (\mathbf{x})\in \lbrack 1,1+d],\ \forall \mathbf{x}\in %
\bR^{3},\ \epsilon (\mathbf{x})\equiv 1,\ \forall \mathbf{x}%
\notin \Omega .  \label{eq:fp3}
\end{equation}%
In other words, the medium is assumed to be homogeneous outside of the
domain $\Omega $. Moreover, for theoretical analysis, we assume that the
coefficient $\epsilon \in C^{3}(\bR^{3})$. We also assume
that the source is placed outside of $\Omega $, i.e.~$\mathbf{x}_{0}\notin 
\overline{\Omega }$.

For the convenience of the theoretical analysis, we state the inverse
problem for the case when the data at the entire boundary are given.
However, in our experiments, only back-scattering data are measured. In
section \ref{sec:num}, we explain how do we work with this type of data.

\textbf{CIP}. \textit{Reconstruct the coefficient $\epsilon (\mathbf{x})$
(or equivalently, the refractive index $n(\mathbf{x})$) for $\mathbf{x}\in
\Omega $, given the measured data at $\partial \Omega $ for a single source
position $\mathbf{x}_{0}\notin \bar{\Omega}$. } 
\begin{equation}
u(\mathbf{x},t)=g(\mathbf{x},t),\mathbf{x}\in \partial \Omega ,t\in
(0,\infty ).  \label{eq:mea}
\end{equation}

The function $g(\mathbf{x,}t)$ represents the time dependent measured data
at the boundary of the domain of interest. The assumption of the infinite
time interval in (\ref{eq:mea}) is not a restrictive one, because in our
method we apply the Laplace transform to $g(\mathbf{x,}t)$ with respect to $%
t $. Since the kernel of this transform decays exponentially with respect to 
$t $, then the Laplace transform effectively cuts off to zero values of the
function $g(\mathbf{x,}t)$ for $t$ larger than some value $T$.

Concerning the uniqueness of this CIP, global uniqueness theorems for
multidimensional CIPs with a single measurement are currently known only
under the assumption that at least one of initial conditions does not equal
zero in the entire domain $\overline{\Omega }$. All these theorems were
proven by the method, which was originated in \cite{BukhKlib}; also,
see, e.g. sections 1.10, 1.11 in \cite{B-K:2012} about this method. This technique is
based on Carleman estimates. Since both initial conditions (\ref{eq:fp2})
equal zero in $\overline{\Omega },$ then this method is inapplicable to our
case. However, since we need to solve numerically our CIP anyway, we assume
that uniqueness takes place.

We remark that equation (\ref{eq:fp1}) is invalid if metallic objects are
present in the domain $\Omega $. To deal with this type of targets, we
follow the suggestion of \cite{K-B-K-S-N-F:IP2012}. It was established
numerically in \cite{K-B-K-S-N-F:IP2012} that metals can be modeled as
dielectrics with a high dielectric constant, which is referred
to as the \textquotedblleft appearing\textquotedblright\ dielectric constant
of metals. It is suggested in \cite{K-B-K-S-N-F:IP2012} that this dielectric
constant can be chosen as 
\begin{equation}
\epsilon \left( \text{metals}\right) \in \left[ 10,30\right] .  \label{X}
\end{equation}

\subsection{The globally convergent method}

\label{sec:gca}

The globally convergent method of \cite{B-K:2012} works with the Laplace
transformed data. However, we do not invert the Laplace transform. Let 
\begin{equation}
w(\mathbf{x},s)=\int\limits_{0}^{\infty }u(\mathbf{x},t)e^{-st}dt,
\label{eq:laplacetr}
\end{equation}%
where the positive parameter $s$ is referred to as the \textit{pseudo
frequency}. We assume that $s\geq \underline{s}>0$, where the number $%
\underline{s}$ is large enough, so that the Laplace transforms of $u$ and
its derivatives $D^{k}u,k=1,2,$ converge absolutely. It follows from (\ref%
{eq:fp1}) that the function $w$ satisfies the equation 
\begin{equation}
\Delta w(\mathbf{x},s)-s^{2}\epsilon (\mathbf{x})w(\mathbf{x},s)=-\delta (%
\mathbf{x}-\mathbf{x}_{0}),\mathbf{x}\in \bR^{3},\ s\geq 
\underline{s}.  \label{eq:w}
\end{equation}%
It was proved in Chapter 2 of \cite{B-K:2012} that $w(\mathbf{x},s)>0$ and $%
\lim\limits_{|\mathbf{x}|\rightarrow \infty }w(\mathbf{x},s)=0$. Define the
function $v$, 
\begin{equation}
v:=\frac{\ln w}{s^{2}}.  \label{eq:v}
\end{equation}%
Substituting $w=e^{vs^{2}}$ into (\ref{eq:w}) and keeping in mind that $%
\mathbf{x}_{0}\notin \overline{\Omega }$, we obtain 
\begin{equation}
\Delta v+s^{2}|\nabla v|^{2}=\epsilon (\mathbf{x}),\mathbf{x}\in \Omega .
\label{eq:c}
\end{equation}%
Equation (\ref{eq:c}) shows that the coefficient $\epsilon (\mathbf{x})$ can
be computed directly via the function $v$. To compute $v$, we eliminate the
unknown coefficient $\epsilon (\mathbf{x})$ from (\ref{eq:c}) by taking the
derivative with respect to $s$ both sides of (\ref{eq:c}). Denote by $%
q:=\partial _{s}v$. Then 
\begin{equation}
v=-\int\limits_{s}^{\infty }qd\tau =-\int\limits_{s}^{\bar{s}}qd\tau +V,
\label{eq:v2}
\end{equation}%
where $\bar{s}>\underline{s}$ is a large number and $V(\mathbf{x}):=v(%
\mathbf{x},\bar{s})$. The number $\bar{s}$ plays the role of a
regularization parameter in the globally convergent method and it is chosen
numerically in the computational practice. The function $V(\mathbf{x}%
):=v\left( x,\overline{s}\right) $ is called the \textquotedblleft tail
function\textquotedblright . It follows from (\ref{eq:v}) and (\ref{eq:v2})
that 
\begin{equation}
V(\mathbf{x})=\frac{\ln w(\mathbf{x},\bar{s})}{\bar{s}^{2}}.  \label{eq:tail}
\end{equation}%
From (\ref{eq:c}) we obtain the following differential integral equation
involving $q$ and $V$ 
\begin{eqnarray}
\Delta q &&-2s^{2}\nabla q\cdot \int\limits_{s}^{\bar{s}}\nabla q(\mathbf{x}%
,\tau )d\tau +2s^{2}\nabla V\cdot \nabla q+2s\left\vert \int\limits_{s}^{%
\bar{s}}\nabla q(\mathbf{x},\tau )d\tau \right\vert ^{2}  \notag \\
&&-4s\nabla V\cdot \int\limits_{s}^{\bar{s}}\nabla q(\mathbf{x},\tau )d\tau
+2s\left\vert \nabla V\right\vert ^{2}=0,\mathbf{x}\in \Omega .  \label{eq:q}
\end{eqnarray}%
Equation (\ref{eq:q}) is coupled with the following Dirichlet boundary
condition for $q$, which follows from (\ref{eq:mea}), 
\begin{equation}
q(\mathbf{x},s)=\psi (\mathbf{x},s),\mathbf{x}\in \partial \Omega ,
\label{eq:q2}
\end{equation}%
where 
\begin{equation}
\psi (\mathbf{x},s)=\frac{\partial }{\partial s}\left[ \frac{\ln (\varphi )}{%
s^{2}}\right] =\frac{\partial _{s}\varphi }{s^{2}\varphi }-\frac{2\ln
(\varphi )}{s^{3}}.  \label{eq:funpsi}
\end{equation}%
Here $\varphi $ is the Laplace transform of the measured data, i.e. $\varphi
(\mathbf{x},s)=\int\limits_{0}^{\infty }g(\mathbf{x},t)e^{-st}dt$.

Note that equation (\ref{eq:q}) has two unknown functions $q$ and $V$. In
order to approximate both of them we treat them differently. In particular,
we use an iterative procedure with respect to the tail function $V$ as
described below.

\subsection{Discretization and description of the algorithm}\label{subsec:alg}

Divide the pseudo frequency interval $[\underline{s},\bar{s}]$ into $N$
uniform sub-intervals by $\bar{s}=s_{0}>s_{1}>\cdots >s_{N}=\underline{s},\
s_{n}-s_{n+1}=h.$ We approximate $q$ by $q(\mathbf{x},s)\approx q_{n}(%
\mathbf{x}),\ s\in (s_{n},s_{n-1}],n=1,\dots ,N.$ We also set $q_{0}\equiv 0$%
. Then after some manipulations, a system of elliptic equations for
functions $q_{n}\left( x\right) $ is derived from (\ref{eq:q}) using the
so-called \textquotedblleft Carleman Weight Function" $\exp \left[ \lambda
\left( s-s_{n-1}\right) \right] ,s\in \left( s_{n},s_{n-1}\right) ,$ where $%
\lambda >>1$ is a certain parameter. We take $\lambda =20$ in all our
computations. This system is 
\begin{eqnarray}
\Delta q_{n} &+&A_{1,n}\nabla q_{n}\cdot \left( \nabla V_{n}-\nabla 
\overline{q_{n-1}}\right)  \notag \\
&=&A_{2,n}|\nabla q_{n}|^{2}+A_{3,n}\left( |\nabla \overline{q_{n-1}}%
|^{2}+|\nabla V_{n}|^{2}-2\nabla V_{n}\cdot \nabla \overline{q_{n-1}}\right)
,  \label{eq:q5}
\end{eqnarray}%
where $A_{i,n},i=1,2,3,$ are some coefficients, depending on $s_{n}$ and $%
\lambda $, which are analytically computed, and $\nabla \overline{q_{n-1}}%
=h\sum_{j=0}^{n-1}\nabla q_{j}$. Here we indicate the dependence of the tail
function $V:=V_{n}$ on the number $n$, because we approximate $V$
iteratively. The discretized version of the boundary condition (\ref{eq:q2})
is given by 
\begin{equation}
q_{n}(\mathbf{x})=\psi _{n}(\mathbf{x}):=\frac{1}{h}\int%
\limits_{s_{n}}^{s_{n-1}}\psi (\mathbf{x},s)ds\approx \frac{1}{2}[\psi (%
\mathbf{x},s_{n})+\psi (\mathbf{x},s_{n-1})],\mathbf{x}\in \partial \Omega .
\label{eq:q7}
\end{equation}%
One can prove that $\left\vert A_{2,n}\right\vert \leq C/\lambda $ for
sufficiently large $\lambda ,$ where $C>0$ is a certain constant. Hence, the
first term on the right hand side of (\ref{eq:q5}) is dominated by the other
terms. Therefore, in the following we set $A_{2,n}|\nabla q_{n}|^{2}:=0$ and
ignore this term in our computations. The system of elliptic equations (\ref%
{eq:q5}) with boundary conditions (\ref{eq:q7}) can be solved sequentially
starting from $n=1$. To solve it, we make use of the iterative process: For
a given $n$ and some approximation $q_{n,i-1}$ of $q_{n}$, we find the next
approximation $q_{n,i}$ of $q_{n}$ by solving (\ref{eq:q5})--(\ref{eq:q7}).
Denote by $m_{n}$ the number of these iterations. The full algorithm is
described as follows.

\subsubsection*{Globally convergent algorithm}

\begin{itemize}
\item Given the first tail $V_0$. Set $q_0 \equiv 0$.

\item For $n = 1, 2, \dots, N$

\begin{enumerate}
\item Set $q_{n,0} = q_{n-1}$, $V_{n,1} = V_{n-1}$

\item For $i = 1, 2,\dots, m_n$

\begin{itemize}
\item Find $q_{n,i}$ by solving (\ref{eq:q5})--(\ref{eq:q7}) with $%
V_{n}:=V_{n,i}$.

\item Compute $v_{n,i} = -h q_{n,i} - \overline{q_{n-1}} + V_{n,i}, \mathbf{x%
} \in \Omega$.

\item Compute $\epsilon _{n,i}$ via (\ref{eq:c}). Then solve the forward
problem (\ref{eq:fp1})--(\ref{eq:fp2}) with the new computed coefficient $%
\epsilon :=\epsilon _{n,i}$, compute $w:=w_{n,i}$ by (\ref{eq:laplacetr})
and update the tail $V_{n,i+1}$ by (\ref{eq:tail}).
\end{itemize}

End (for $i$)

\item Set $q_{n}=q_{n,m_{n}}$, $\epsilon _{n}=\epsilon
_{n,m_{n}},V_{n}=V_{n,m_{n}+1}$ and go to the next frequency interval $\left[
s_{n+1},s_{n}\right] $ if $n<N.$ If $n=N$, then stop.
\end{enumerate}
\end{itemize}

Stopping criteria of this algorithm with respect to $i,n$\ should be
addressed computationally, see details in section \ref{sec:num}.

\subsection{The initial tail function and the global convergence}

\label{sec:gc}

We remark that the convergence of this algorithm depends on the choice of
the initial tail function $V_{0}$. In \cite{B-K:2012}, see also \cite%
{B-K:JIIP2012}, the global convergence of this algorithm was proved within
the context of an approximate mathematical model. First, it has been proved
in \cite{B-K:2012} that under some conditions, there exists a function $p(%
\mathbf{x})\in C^{2+\alpha }\left( \overline{\Omega }\right) $ such that 
\begin{equation}
V(\mathbf{x},s)=\frac{p(\mathbf{x})}{s}+O\left( \frac{1}{s^{2}}\right)
,s\rightarrow \infty ,  \label{3.13}
\end{equation}%
where $C^{2+\alpha }\left( \overline{\Omega }\right) $ is the H\"{o}lder
space. Due to this asymptotic behavior, we assume that the exact tail is
given by 
\begin{equation}
V(\mathbf{x},s)=\frac{p\left( \mathbf{x}\right) }{s}=\frac{\ln w\left( 
\mathbf{x},s\right) }{s^{2}},\forall s\geq \overline{s}.  \label{eq:model2}
\end{equation}%
We use this assumption only on the initializing iteration to obtain $%
V_{0}\left( \mathbf{x}\right) .$ That is, we make use of only the first
order term in the asymptotic expansion (\ref{3.13}). Under this assumption,
it follows from (\ref{eq:q}), (\ref{eq:q2}) that $p\left( \mathbf{x}\right) $
satisfies 
\begin{eqnarray}
\Delta p(\mathbf{x}) &=&0,\mathbf{x}\in \Omega ,\ p\in C^{2+\alpha }\left( 
\overline{\Omega }\right) ,  \label{3.37} \\
p|_{\partial \Omega } &=&-\overline{s}^{2}\psi \left( \mathbf{x},\overline{s}%
\right) ,  \label{3.38}
\end{eqnarray}%
As the first guess for the tail function we take 
\begin{equation}
V_{0}\left( \mathbf{x}\right) :=\frac{p\left( \mathbf{x}\right) }{\overline{s%
}},\mathbf{x}\in \Omega .  \label{3.43}
\end{equation}%
With this choice of the initial tail function, it was proved in Chapter 2 of 
\cite{B-K:2012}, Theorem 2.9.4 (see also Theorem 5.1 in \cite{B-K:JIIP2012})
that the proposed algorithm is convergent in the following sense: it
delivers some points in a sufficiently small neighborhood of the exact
coefficient $\epsilon (\mathbf{x})$. The latter is sufficient for
computational purposes. The size of this neighborhood depends on the noise
in the data $\psi ,$ the discretization step size $h$ of the interval $\left[
\underline{s},\overline{s}\right] $ and the domain $\Omega .$ We note that
no \textit{a priori} information about the unknown coefficient is used here.
Therefore, we say that the algorithm is globally convergent within the
framework of the approximation (\ref{eq:model2}).

In our computations, the bound constraints (\ref{eq:fp3}) is used to truncate the coefficient $\epsilon _{n,i}(\mathbf{x})$ on each
iteration.

\subsection{Numerical solution of the forward problem}

Although a point source is used in the forward model (\ref{eq:fp1})--(\ref%
{eq:fp2}) for theoretical analysis, we make use of an incident plane wave in
our numerical implementation. Moreover, since it is impossible to solve the
forward problem numerically in the infinite space $\bR^{3}$,
we consider the wave equation in a bounded domain $G\subset %
\bR^{3}$ such that $\Omega \subset G$. For simplicity, we
choose $G$ as the prism 
\begin{equation}
G:=\{\mathbf{x}\in \bR^{3}:X_{l}\leq x\leq X_{u},Y_{l}\leq
y\leq Y_{u},Z_{l}\leq z\leq Z_{u}\}.  \notag
\end{equation}%
We denote by $\partial G_{z}^{l}:=\{z=Z_{l}\}$, $\partial
G_{z}^{u}:=\{z=Z_{u}\}$ and $\partial G_{xy}=\partial G\setminus (\partial
G_{z}^{l}\cup \partial G_{z}^{u})$. An incident plane wave of a short time
period is excited at $\partial G_{z}^{u}$ and propagates in the negative $z$
direction. At the plane $\partial G_{z}^{l}$ we assume that the absorbing
boundary condition is satisfied, and at $\partial G_{xy}$ we assign the
homogeneous Neumann boundary condition. More precisely, we solve the
following problem 
\begin{eqnarray}
\epsilon (\mathbf{x})u_{tt}(\mathbf{x},t) &=&\Delta u(\mathbf{x},t),(\mathbf{%
x},t)\in G\times (0,T),  \label{eq:fp01} \\
u(\mathbf{x},0) &=&0,\quad u_{t}(\mathbf{x},0)=0,\mathbf{x}\in G,
\label{eq:fp03} \\
\partial _{\nu }u &=&f(t),(\mathbf{x},t)\in \partial G_{z}^{u}\times
(0,t_{1}),  \label{eq:fp04} \\
\partial _{\nu }u &=&-u_{t},(\mathbf{x},t)\in \partial G_{z}^{u}\times
(t_{1},T),  \label{eq:fp05} \\
\partial _{\nu }u &=&-u_{t},(\mathbf{x},t)\in \partial G_{z}^{l}\times (0,T),
\label{eq:fp06} \\
\partial _{\nu }u &=&0,(\mathbf{x},t)\in \partial G_{xy}\times (0,T),
\label{eq:fp07}
\end{eqnarray}%
where $\nu $ is the outward normal vector of $\partial G$ and $t_{1}:=2\pi
/\omega $ is the duration of the excitation of the incident plane wave.
Function $f$ is the incident waveform chosen by 
\begin{equation}
f(t)=\sin (\omega t),0\leq t\leq t_{1}=2\pi /\omega .  \notag
\label{eq:incfun}
\end{equation}%
Here $\omega $ represents the angular frequency of the incident plane wave.
The forward problem (\ref{eq:fp01})--(\ref{eq:fp07}) is solved using the
software package WavES \cite{waves} via the hybrid FEM/FDM method described
in \cite{BSA}.

\section{The experimental setup and data pre-processing}

\label{sec:exp}

\subsection{Data acquisition}

\begin{figure}[tph]
\centering
\subfloat[]{\includegraphics[width = 0.45\textwidth,height=0.3\textwidth]{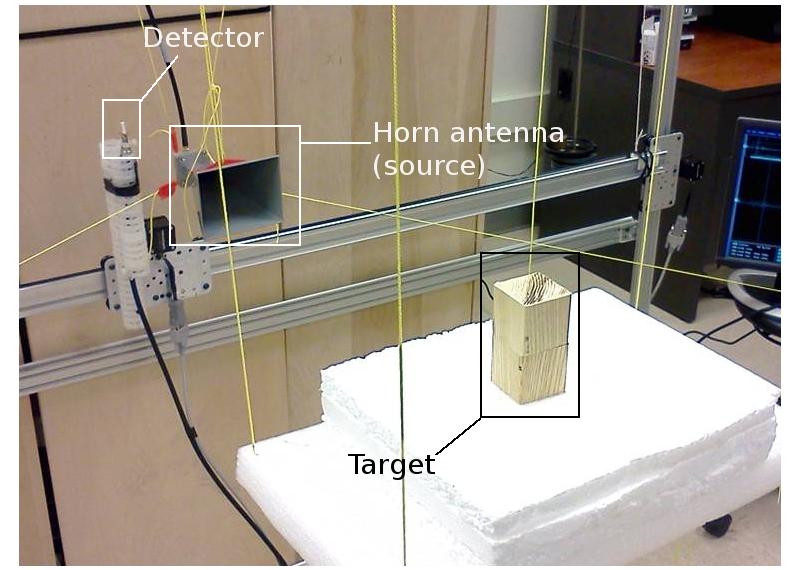}} \hspace{%
0.3truecm} 
\subfloat[]{\includegraphics[width =
0.37\textwidth,height=0.3\textwidth]{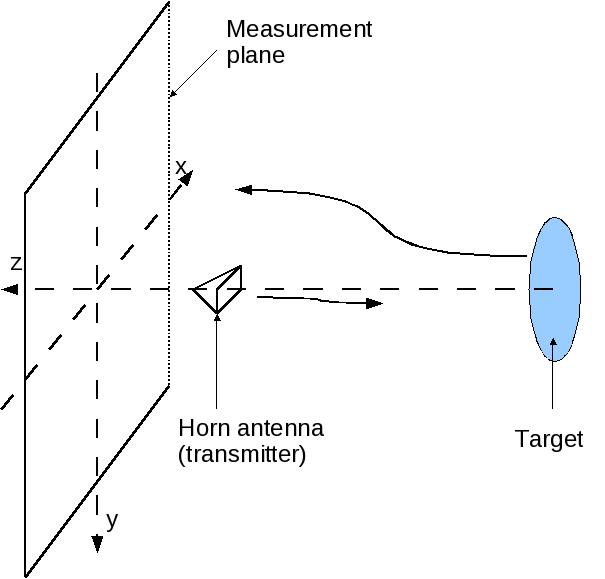}}
\caption{(a): A picture of our experiment setup; (b) Diagram of our setup.}
\label{fig:setup}
\end{figure}

Our experimental configuration is shown in Figure~\ref{fig:setup}. The setup
of our measurements included a horn antenna (transmitter) fixed at a given
position and a detector scanned in a square of a vertical plane, which we
refer to as the measurement plane. Consider the Cartesian coordinate system $%
Oxyz$ as shown in Figure \ref{fig:setup}(b). The scanning area was 1 m by 1
m with the step size of 0.02 m, starting at $(x,y)=(-0.5,-0.5)$, and ends at 
$(x,y)=(0.5,0.5)$. 

In our model (\ref{eq:fp1})--(\ref{eq:fp2}), we assumed that the source
point $\mathbf{x}_{0}$ is in $\bR^{3}\setminus \overline{%
\Omega }$. However, due to some technical difficulties with the mechanical
scanning system, the horn antenna was not placed behind but in front of the
measurement plane (between the measurement plane and the targets). Therefore
a small area in the center of the scanning area on the measurement plane was
shaded by the horn. The horn was placed at the distance of about 0.2-0.25 m
from the measurement plane and the distances from the targets to the
measurement plane are about 0.8 m.

At each position of the detector, a number of electric pulses were emitted
by the horn. The detector received two types of signals: the direct signals
from the source and the backscattered signals by the targets. The direct
signals are used for time reference in data pre-processing. There were also
other unwanted signals due to scattering by some objects in the room. To
reduce the instability of the recorded signals in terms of magnitude, the
measurements were repeated 800 times at each detector position and the
recorded signals were averaged. By scanning the detector and repeating the
measurements, we obtained essentially the same signals as using one incident
signal and multiple detectors at the same time.

Pulses were generated by the Picosecond Pulse Generator 10070A. The
scattered signals were measured by a Tektronix DSA70000 series real-time
oscilloscope. The emitted pulses were of 300 picoseconds duration. The
wavelength of the incident pulses was about 0.03 m. The sampling rate (the
step size in time between two consecutive records of captured signals) was $%
\Delta t=10$ picoseconds. Each signal was recorded for 10 nanoseconds.

\subsection{Data pre-processing}

\begin{figure}[tph]
\centering
\subfloat[]{\includegraphics[width =
0.7\textwidth,height=0.38\textwidth]{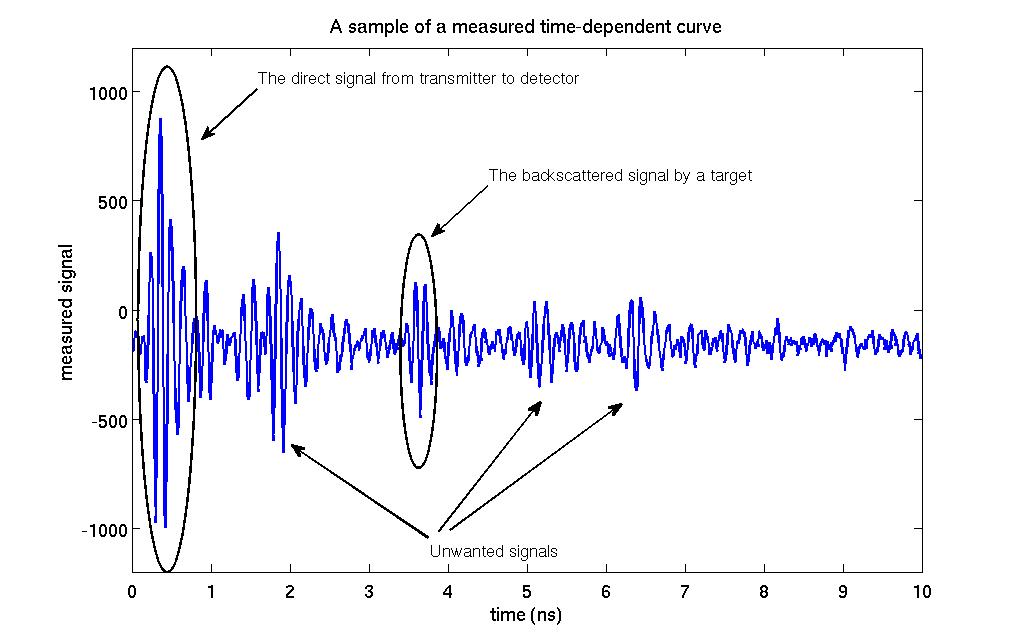}}
\par
\subfloat[]{\includegraphics[width =
0.47\textwidth,height=0.34\textwidth]{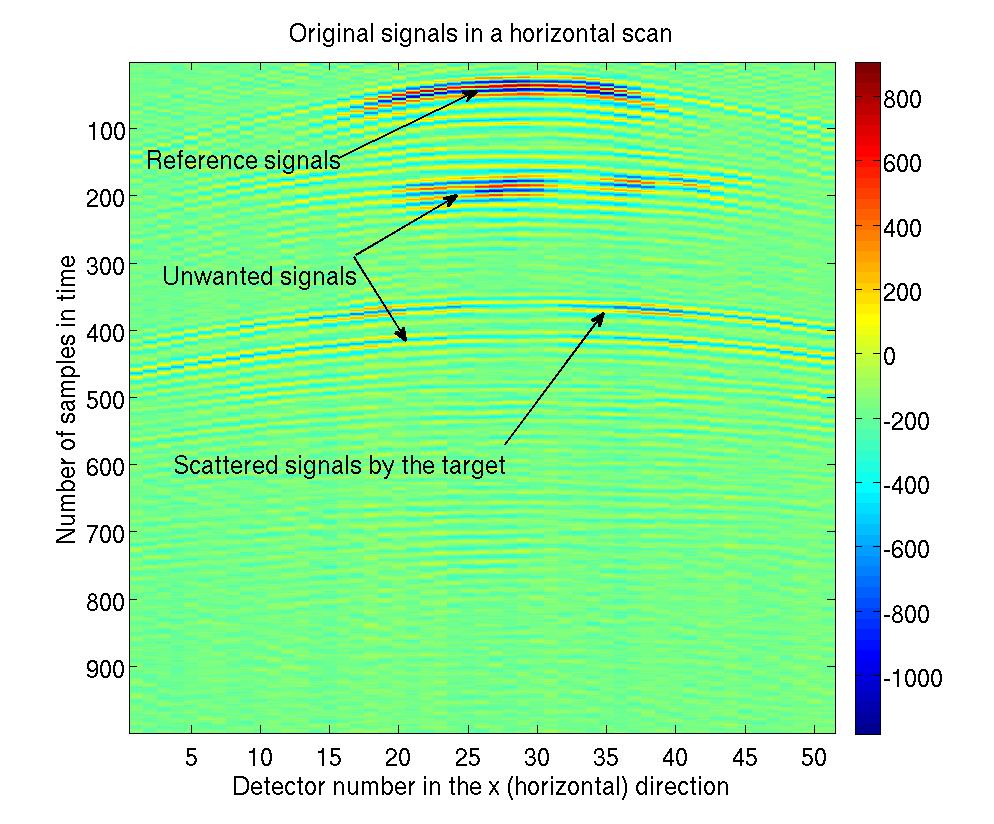}} 
\subfloat[]{\includegraphics[width =
0.47\textwidth,height=0.34\textwidth]{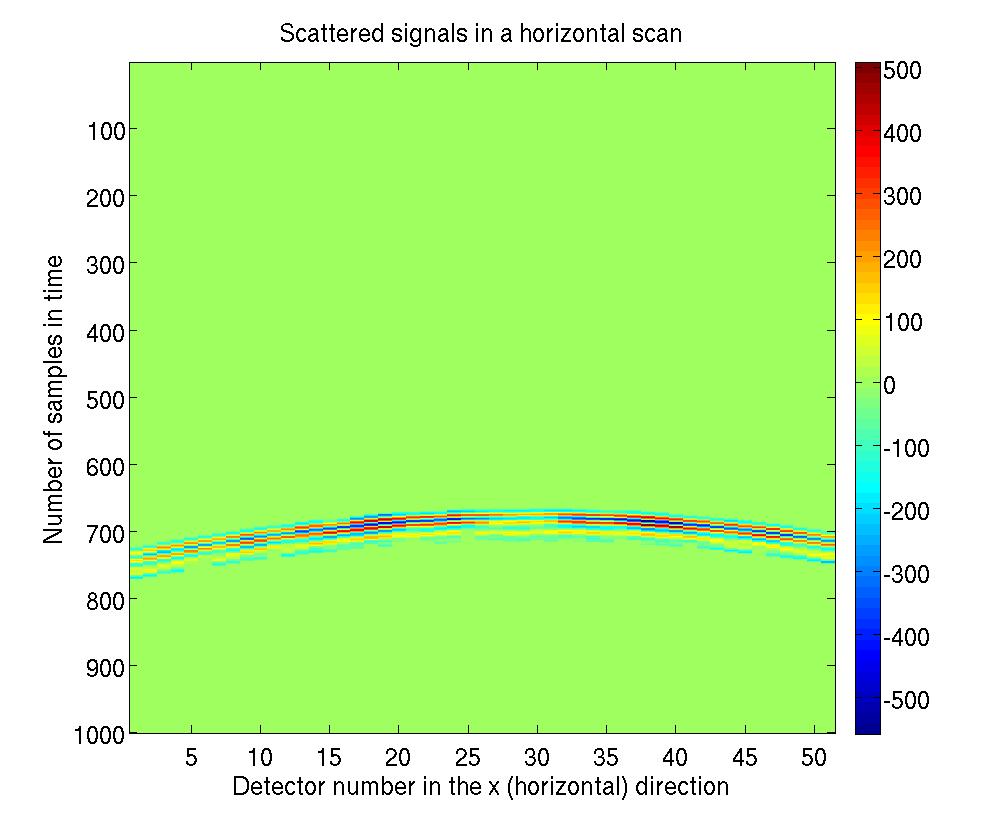}}
\caption{An experimental data set: (a) a 1D curve; (b) a 2D data on a
horizontal scan; (c) After steps 1-4 of the data pre-processing.}
\label{fig:prep02}
\end{figure}

\begin{figure}[tph]
\centering
\subfloat[]{\includegraphics[width = 0.45\textwidth,height=0.32\textwidth]{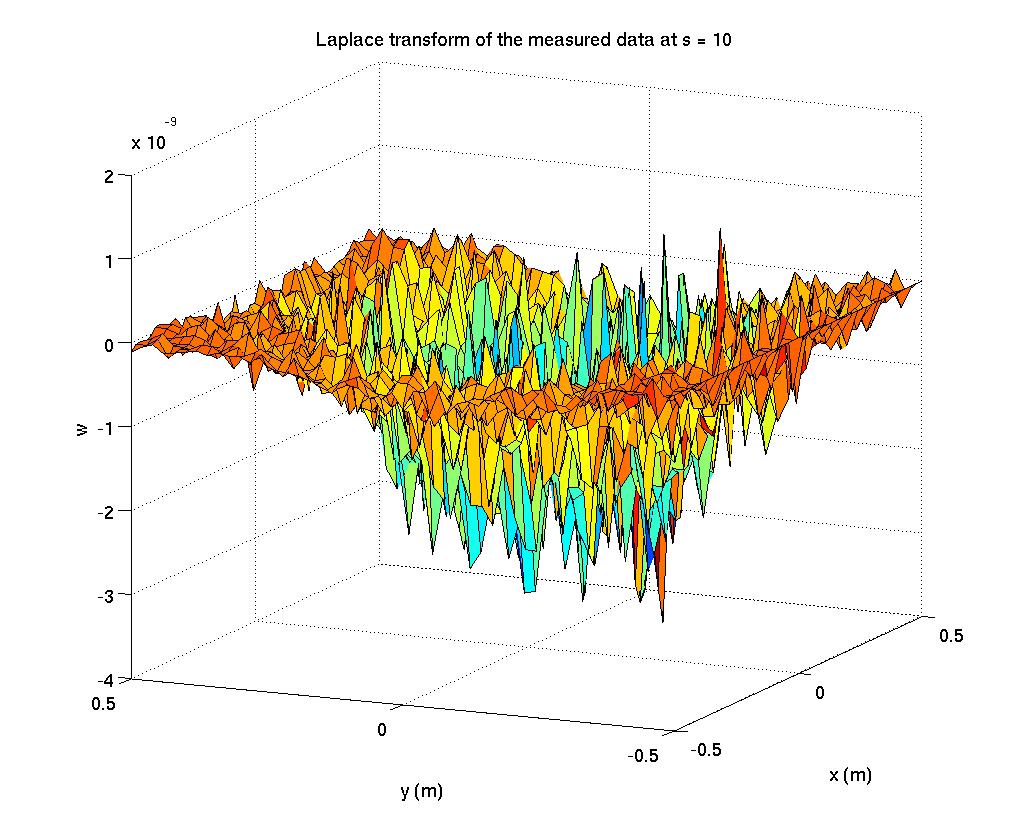}}
\subfloat[]{\includegraphics[width = 0.45\textwidth,height=0.32\textwidth]{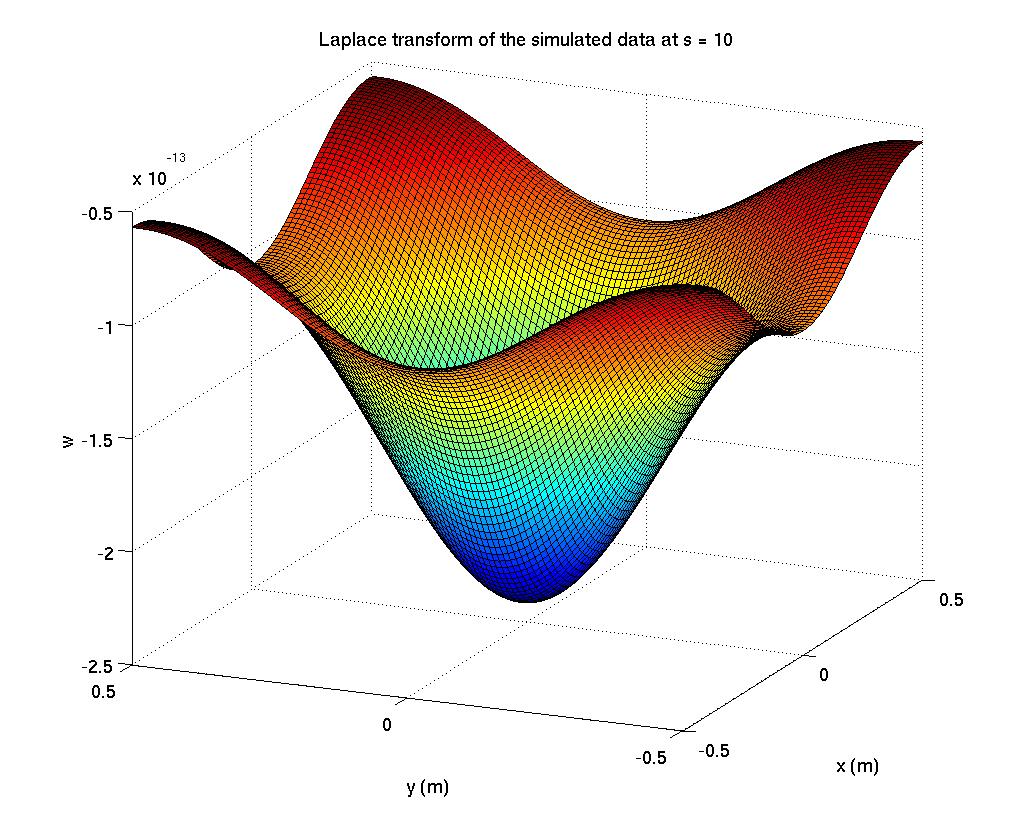}}
\caption{Experimental vs.~simulated scattered waves on the measurement plane
after Laplace transform: (a) measured data; b) simulated data.}
\label{fig:prep1}
\end{figure}

One of the biggest challenges in working with these experimental data is to deal
with the \textit{huge misfit} between these data and the data produced via
computational simulations, also see \cite%
{B-K:2012,B-K:IP2010,K-F-B-P-S:IP2010,K-B-K-S-N-F:IP2012} for the same
conclusion. There are several causes of this misfit such as (i) the
instability of the amplitude of the emitted signals (incident waves) which
causes the instability of the received signals, (ii) unwanted scattered
waves by several existing objects around our devices (see Figure~\ref%
{fig:prep02}(a)), (iii) the shadow on the measurement plane caused by the
transmitting horn antenna, and (iv) the difference between the experimental
and simulated incident waves. Figure \ref{fig:prep1} compares the Laplace
transform of an experimental scattered wave and the corresponding simulated
one, which shows a huge misfit between them. Note that the Laplace transform
of the experimental wave was carried out after removing the incident wave
and unwanted parts, as shown in Figure \ref{fig:prep02}(c).

Therefore, the \emph{central procedure} before applying inversion methods is
the data pre-processing. This procedure is usually heuristic and cannot be
rigorously justified. Our data pre-processing consists of six steps
described below. We do not describe steps 1-3 in detail here, since they are
straightforward.

\noindent \textit{Step 1. Off-set correction}. The acquired signals are
usually shifted from the zero mean value. This can be corrected by
subtracting the mean value from them.

\noindent \textit{Step 2. Time-zero correction}. Time-zero refers to the
moment at which the signal is emitted from the transmitter. The recorded
signals may be shifted in time. We use the direct signals from the
transmitter to the detector to correct the time-zero.

\noindent \textit{Step 3. Source shift}. As mentioned above, the horn antenna in our experiments is placed between the targets and the measurement
plane. However, in data calibration, we need to simulate the data for the case when the measurement plane is between the horn and the targets. Therefore, we artificially
\textquotedblleft shift\textquotedblright\ the horn in the positive $z-$%
direction such that it is 0.4 m further than the measurement plane from the
targets. It is done by shifting the whole time-dependent data by a number of
samples which corresponds to the shifted distance.

\noindent \textit{Step 4. Extraction of scattered signals}. Apart from the
signals backscattered by the targets, our measured data also contain various
types of signals as mentioned above. What we need, however, is the scattered
signals by the targets only. The extraction of these scattered signals for
each target is done as follows. First, we exclude the direct signals and the
unwanted signals, which come earlier than the scattered signals by the
target (see Figure \ref{fig:prep02}(a)-(b)), by calculating the time of
arrival. These unwanted signals are due to the reflection of the direct
signals by the metallic structure behind the measurement plane, so we can
estimate their times of arrival as we know the distance from the measurement
plane to this structure. Then, as we observed that the scattered signals by
the target are the strongest peaks of the remaining, we first detect, for
each detector position, the strongest negative peak after removing the
aforementioned signals. Then, the scattered signals by the target are taken
as 7 peaks (4 negative peaks and 3 positive peaks) starting from the 
first negative peak prior to the strongest one, if its amplitude is less
than 80\% of the strongest one (see Figure \ref{fig:prep02}(a)). Otherwise,
we start from the second negative peak prior to the strongest one. The
reason for choosing 7 peaks for the scattered signals is due to the fact
that the incident pulses also contain 7 strong peaks. We note that having
the scattered signals by the target, we can easily determine the distance
from its front side to the measurement plane by calculating the time of
arrival.

\noindent \textit{Step 5. Data propagation}. After getting the scattered
signals, the next step of data pre-processing is to propagate the data
closer to the targets, i.e.~to approximate the scattered waves on a plane
closer to the targets, compared to the measurement plane. There are two
reasons for doing this. The first one is that since the Laplace transform
decays exponentially in terms of the time delay, which is proportional to
the distance from the targets to the measurement plane, then the amplitude
of the data after the Laplace transform on the measurement plane is very
small and can be dominated by the computational error. The second reason is
that this propagation procedure helps to reduce the computational cost
substantially as the computational domain $\Omega $ is reduced. We have also
observed that the data propagation helps to reduce the noise in the measured
data.

\noindent \textit{Step 6. Data calibration}. Finally, since the amplitude of
the experimental incident and scattered waves are usually quite different
from simulations, we need to bring the former to the same level of the
amplitude as the latter. This is done using a known target referred to as 
\textit{calibrating object}.

Figure \ref{fig:prep02} shows an original data set and data after steps 1-4
for a horizontal scan of the detector. One can see that steps 1-4 help to
remove the unwanted signals. In the following, we present our methods for
steps 5 and 6.

\subsubsection{Data propagation}

\label{subsec:datprop}

Denote by $P_{m}$ the measurement plane and by $P_{p}$ the propagation
plane, which is closer to the targets than $P_{m}$. Without a loss of the
generality, we denote by $P_{m}=\{z=a\}$ and $P_{p}=\{z=0\}$, where the
number $a>0.$ Moreover, denote by $\Gamma =(-0.5,0.5)\times
(-0.5,0.5)\}\subset \bR^{2}$ the scanning area of the
detector on the plane $P_{m}$. Let $\Gamma _{m}=\{(x,y,a)\in %
\bR^{3}:(x,y)\in \Gamma \}$ and $\Gamma _{p}=\{(x,y,0)\in %
\bR^{3}:(x,y)\in \Gamma \}$. We also denote by $u^{s}(%
\mathbf{x},t)$ the scattered wave. Note that the medium between $P_{m}$ and $%
P_{p}$ is homogeneous with $\epsilon =1$ and the scattered wave $u^{s}$
propagates in the direction from $P_{p}$ to $P_{m}$. The aim of the data
propagation is to approximate $u^{s}\big\vert_{\Gamma _{p}\times (0,\infty
)} $ from the measured data $\widetilde{g}(x,y,t):=u^{s}\big\vert_{\Gamma
_{m}\times (0,\infty )}$.

To do this, we make use of a time reversal method. Its idea is to reverse
the scattered wave in time via solution of an initial boundary value problem
for the time-reversed wave function. We proceed as follows. %

Since short pulses are used as incident waves, it is reasonable to assume
that the scattered wave $u^{s}$ in the domain between $P_{m}$ and $P_{p}$
vanishes along with its time derivative $u_{t}^{s}$ after some time $T$.
Therefore in the following we only consider the finite time interval $(0,T)$%
. Denote $\tau :=T-t$. Then the time-reversed wave function $u^{r}(\mathbf{x}%
,\tau ):=u^{s}(\mathbf{x},t)$ satisfies the homogeneous wave equation.
Moreover, it propagates in the negative $z$ direction, i.e.~from $P_{m}$ to $%
P_{p}$. To find $u^{r}\big\vert_{\Gamma _{p}}$, we consider the domain $D:=\{%
\mathbf{x}\in \bR^{3}:(x,y)\in \Gamma ,b<z<a\}$ with $b<0$.
Note that $\Gamma _{p}\subset D.$ The reason for choosing this larger domain
is that we need to assign boundary conditions at $\partial D$. Indeed, we
assume that $u^{r}$ satisfies the absorbing boundary condition at $\Gamma
_{b}:=\{(x,y,b):(x,y)\in \Gamma \}$. On $\Gamma _{b},$ far from our
propagation plane, this boundary condition means, heuristically, that we
\textquotedblleft send back" the original scattered wave $u^{s}$ recorded at 
$P_{m}$. On the other hand, we impose the zero Neumann boundary condition at
the rest of the boundary of $D$, except of $\Gamma _{m}$. Denote $%
Q_{T}=D\times (0,T)$ and $\Gamma _{3}:=\partial D\setminus (\Gamma _{m}\cup
\Gamma _{b})$. We obtain the following problem for
the function $u^{r}(\mathbf{x},\tau )$ 
\begin{eqnarray}
&&u_{\tau \tau }^{r}=\Delta u^{r},(\mathbf{x},\tau )\in Q_{T},
\label{eq:datpro2} \\
&&u^{r}\left( \mathbf{x},0\right) =u_{\tau }^{r}\left( \mathbf{x},0\right)
=0,\mathbf{x}\in D,  \label{eq:datpro3} \\
&&u^{r}\big\vert_{\Gamma _{m}\times (0,T)}=\widetilde{g}(x,y,T-\tau ),
\label{eq:datpro4} \\
&&\left( \partial _{\nu }u^{r}+\partial _{\tau }u^{r}\right) \big\vert%
_{\Gamma _{b}\times (0,T)}=0,  \label{eq:datpro5} \\
&&\partial _{\nu }u^{r}\big\vert_{\Gamma _{3}\times (0,T)}=0.
\label{eq:datpro6}
\end{eqnarray}%
A similar procedure was proposed and numerically implemented for
computationally simulated data in \cite{AKN:WM2012}. However,
the absorbing boundary condition for the original scattered wave $u^{s}$ was
applied in \cite{AKN:WM2012}. Since our time-reversed wave
function $u^{r}$ propagates from $\Gamma _{m}$ to $\Gamma _{b}$, we believe
that it is more natural to apply the absorbing boundary condition on $\Gamma
_{b}$ to $u^{r}$. Note that we do not locate scatterers here, since we know
from our previous pre-processing steps an approximate distance to the
targets. We refer the reader to \cite{Fink:PT1997,LRTF:Sci2007} and the
references therein for experimental time reversal mirror method, a similar
procedure but using a physical mirror to reverse the received signals in
time and send back to the targets.

Theorem \ref{th:datpro} below shows the stability of the problem (\ref%
{eq:datpro2})--(\ref{eq:datpro6}). We do not prove this theorem here due to
the space limit. This theorem can be extended to more general domains. We
note that its proof is similar with the proof of Theorem 4.1 of \cite%
{Beilina:CEJM2013} for the Maxwell's system. For brevity we are not
concerned here with minimal smoothness assumptions and leave aside the
question of existence. We conjecture that it can be addressed via the
technique of chapter 4 of \cite{Ladyzhenskaya:1985}.

\begin{theorem}
\label{th:datpro} Assume that there exists a solution $u^{r}\in H^{2}\left(
Q_{T}\right) $ of the problem (\ref{eq:datpro2})-(\ref{eq:datpro6}). Also,
assume that the function $\widetilde{g}\in H^{2}\left( \Gamma _{m}\times
\left( 0,T\right) \right) $ and there exists such a function $F\in
H^{2}\left( Q_{T}\right) $ that 
\begin{eqnarray}
&&F\left( \mathbf{x},0\right) =F_{\tau }\left( \mathbf{x},0\right) =0,\left(
\partial _{\nu }F+F_{\tau }\right) \mid _{\Gamma _{b}\times \left(
0,T\right) }=0,\partial _{\nu }F\mid _{\Gamma _{3}\times \left( 0,T\right)
}=0,  \notag \\
&& F\mid _{\Gamma _{m}\times \left( 0,T\right) } =\widetilde{g}\left( \mathbf{%
x},t\right) ,  \quad
\left\Vert F\right\Vert _{H^{2}\left( Q_{T}\right) } \leq C\left\Vert 
\widetilde{g}\right\Vert _{H^{2}\left( \Gamma _{m}\times \left( 0,T\right)
\right) },  \notag
\end{eqnarray}%
where $C>0$ is a certain number. Then that solution $u^{r}$ is unique and
the following stability estimate holds with a constant $C_{1}=C_{1}\left(
C,Q_{T}\right) >0$ depending only on the listed parameters 
\begin{equation}
\left\Vert u^{r}\right\Vert _{H^{1}\left( Q_{T}\right) }\leq C_{1}\left\Vert 
\widetilde{g}\right\Vert _{H^{2}\left( \Gamma _{m}\times \left( 0,T\right)
\right) }.  \notag
\end{equation}
\end{theorem}

By solving (\ref{eq:datpro2})--(\ref{eq:datpro6}), we obtain an
approximation of $u^{r}\left( \mathbf{x},\tau \right) $ and then obtain an
approximation of $u^{s}$ for $\mathbf{x}\in \Gamma _{p}$. In this work, we
use the finite difference method to solve this problem. Since the
formulation is very standard, we omit the details here. We note that other
methods can also be used to solve this data propagation problem such as the
Fourier transform method or the quasi-reversibility method. We will discuss
these two methods in future works.

%
%

\subsubsection{Data calibration}

\begin{figure}[tph]
\centering
\subfloat[]{\includegraphics[width =
0.45\textwidth]{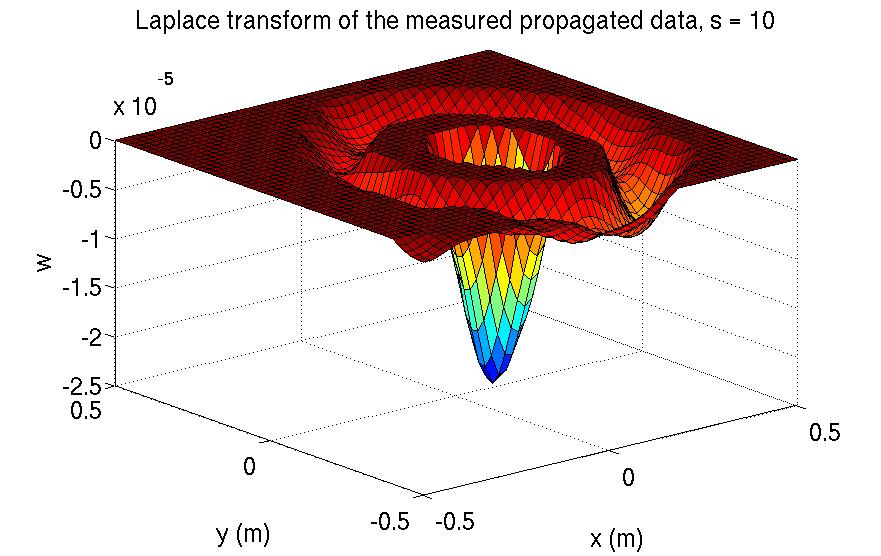}} \subfloat[]{%
\includegraphics[width = 0.45\textwidth]{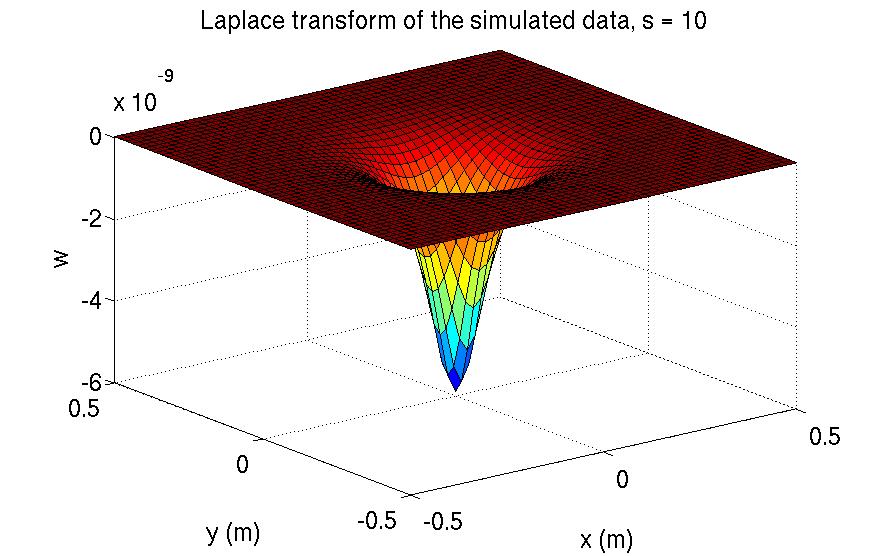}}
\caption{Laplace transform of the scattered wave on the propagation plane $%
P_p$: (a) measured data; b) simulated data.}
\label{fig:prep2}
\end{figure}

Usually the experimental data have quite different amplitudes compared to
the simulations, see Figure \ref{fig:prep2} which shows that the minimum of
the Laplace transform of the propagated measured data is approximately $%
-2\times 10^{-5},$ whereas the minimum of the simulated data is about $%
-5\times 10^{-9}$. We choose a number, which is called \emph{calibration
factor,} to scale the measured data to the same scaling as in our
simulations. To do this, we make use of the measured data of a single
calibrating object whose location, shape, size and material are known. The
word \textquotedblleft single" is important here to ensure that the
calibration procedure is unbiased, i.e. it remains the same for all targets.

First, we computationally simulate the data on $\Gamma _{p}$ for the
calibrating object by solving problem the (\ref{eq:fp01})--(\ref{eq:fp07}).
Next, we compute the Laplace transform (\ref{eq:laplacetr}) of this
computationally simulated solution. Just as in \cite{B-K:2012}, in studies
below we work with $s\in \left[ \underline{s},\overline{s}\right] .$ Numbers 
$\underline{s},\overline{s}$ are chosen numerically, see section \ref%
{sec:num}. Denote by $w_{sim}^{t}\left( \mathbf{x},s\right) $, $%
w_{sim}^{s}\left( \mathbf{x},s\right) $ and $w_{sim}^{i}\left( \mathbf{x}%
,s\right) $ respectively the Laplace transforms of the total wave, the
scattered wave and the incident wave of the simulated solution for the
calibrating object. Clearly, $w_{sim}^{s}\left( \mathbf{x},s\right)
=w_{sim}^{t}\left( \mathbf{x},s\right) -w_{sim}^{i}\left( \mathbf{x}%
,s\right) $. It can be proved that $w_{sim}^{s}\left( \mathbf{x},s\right)
\leq 0$, see Figure \ref{fig:prep2}(b). Figure \ref{fig:prep2}(b) displays
the function $w_{sim}^{s}\left( \mathbf{x},s\right) $ for $\mathbf{x}\in
\Gamma _{p}$ and qualitatively this is a typical behavior for all targets.
Let 
\begin{equation}
d_{sim,s}=\min_{\Gamma _{p}}w_{sim}^{s}\left( \mathbf{x},s\right) .  \notag
\end{equation}%
Next, for $\mathbf{x}\in \Gamma _{p}$ let $w_{exp}^{s}\left( \mathbf{x}%
,s\right) $ be the Laplace transform of the propagated experimental data for
the calibrating object, see Figure \ref{fig:prep2}(a). Denote by 
\begin{equation}
d_{exp,s}=\min_{\Gamma _{p}}w_{exp}^{s}\left( \mathbf{x},s\right) .  \notag
\end{equation}%
The number $d_{sim,s}/d_{exp,s}$ is used as the calibration factor for all
targets at pseudo-frequency $s$. That means, the propagated measured data of
all targets are multiplied by this calibration factor before being used in
the inversion algorithm.

For the data sets used in this paper, we have two types of targets:
dielectric and metallic targets. We have observed that two different
calibration factors should be used for these two types of targets, because
the signals from them have different levels of amplitude. First of all, we
differentiated these two types of targets by comparing the amplitudes of the
recorded signals. Indeed, we have consistently observed that the maximal
values of amplitudes of measured signals are at least two times larger for
metallic targets than for dielectric ones on those positions of detectors
which are most sensitive to the presence of targets. Next, we chose in each
type a known object as the calibrating object. In other words, we should use
a dielectric calibrating object for all dielectric targets and another metal
calibrating object for all metallic targets.

The value of $\epsilon \left( \mathbf{x}\right) $ for the dielectric
calibrating object was taken as $\epsilon \left( \mathbf{x}\right) =4.28$
inside that target and $\epsilon \left( \mathbf{x}\right) =1$ outside of it.
For the metallic calibrating object, as suggested by (\ref{X}), we took $%
\epsilon \left( \mathbf{x}\right) =12$ inside and $\epsilon \left( \mathbf{x}%
\right) =1$ outside of it.

\section{Numerical implementation and reconstruction results}

\label{sec:num}

Now we describe some details of the numerical implementation and present
reconstruction results for our experimental data using the globally
convergent algorithm of section \ref{sec:gca}. In our computations, the
frequency of the simulated incident wave was chosen as $\omega =30$.

There were ten data sets tested in this paper. Each data set consisted of
only one target numbered from 1 to 10. Four of them (targets 1 - 4) were
non-blind and six of them (targets 5 - 10) were blind. In our data
pre-processing, we chose target 1 (a piece of wood) and target 3 (a metallic
sphere) as our calibrating objects.

\subsection{Dimensionless variables}

The spatial dimensions in our experiment were given in meters. Since the
scanning step in our measured data was $0.02$ m in both $x$ and $y$
directions, we chose the dimensionless spatial variable $\mathbf{x}^{\prime
} $ to be $\mathbf{x}^{\prime }=\mathbf{x}/1(m)$. Next, to scale the wave
speed to be $1$ in the homogeneous medium, as in our simulations, we chose
the dimensionless time $t^{\prime }$ by $t^{\prime }=0.3t$ where $t$ is the
time given in nanoseconds (ns). The factor $0.3$ is the speed of light in
meters per nanosecond in the free space. For the simplicity of notations, we
use $\mathbf{x}$ and $t$ again to denote the dimensionless variables.

\subsection{Choosing the domains}

Before applying the inversion algorithm, some information about the targets
was in place already from the measured data due to the data pre-processing.
First, we obtained the distance from the targets to the measurement plane.
Second, by the data propagation, the targets' locations in the $xy$ plane
were estimated. Thirdly, we also differentiated between nonmetallic and
metallic targets directly from the measured data based on signal amplitudes.

Given the estimated distances from the targets to the measurement plane,
which were about $0.8$ m, we propagated the measured data from the
measurement plane $P_{m}=\{z=0.8\}$ to the plane $P_{p}=\{z=0.04\}$. This
means that the distance from the front sides of the targets to the
backscattering boundary of our inversion domain was about 0.04 m. The reason
for choosing this distance was due to good reconstruction results we have
obtained for several non-blind targets. The domain $\Omega $ was chosen by 
\begin{equation}
\Omega =\left\{ \mathbf{x}\in (-0.5,0.5)\times (-0.5,0.5)\times
(-0.1,0.04)\right\} .  \label{Y}
\end{equation}%
For solving the forward problem (\ref{eq:fp01})--(\ref{eq:fp07}), using the
hybrid FDM/FEM method, we chose the domain $G$ as 
\begin{equation}
G=\left\{ \mathbf{x}\in (-0.56,0.56)\times (-0.56,0.56)\times
(-0.16,0.1)\right\} .  \notag
\end{equation}%
This domain $G$ was decomposed into two subdomain: $G=\Omega \cup
(G\setminus \Omega )$. We recall that $\epsilon (\mathbf{x})=1$ in $%
G\setminus \Omega $. Therefore, it is only necessary to solve the inverse
problem in $\Omega $. In $\Omega $ we use a FEM mesh with tetrahedral
elements, while in $G\setminus \Omega $ we use a FDM mesh with the mesh size
of $0.02$ by $0.02$ by $0.02$ in Test 1 and $0.01$ by $0.01$ by $0.01$ in
Test 2 below. The reason for using the FEM mesh in $\Omega $ is that it is
possible to refine the reconstruction using adaptive mesh refinement.
However, we do not discuss this step in this work. We refer to \cite%
{B-K:2012, BK:AA2013} for more details about the adaptivity.

The time interval $(0,T)$ in the forward problem (\ref{eq:fp01})--(\ref%
{eq:fp07}) was chosen equal to $(0,1.2)$. Since the explicit scheme in time
was used in WaveES, the time step size was chosen as $\Delta t= 0.0015$
which satisfies the CFL stability condition.

The pseudo frequencies $s_{n}$ were chosen from $\underline{s}=8$ to $%
\overline{s}=10$ with the step size $h=0.05$. This frequency interval was
chosen because it gave good reconstructions of the non-blind targets.

\subsection{Estimation of the $xy$ projection}

\label{subsec:cs}

During our data pre-processing for non-blind targets, we observed that the $%
xy$ projection of a target can be roughly estimated directly from the
propagated data. Indeed, we define $\Gamma _{T}$ as 
\begin{equation}
\Gamma _{T}=\{(x,y):v_{prop}(x,y,\bar{s})<0.85\min v_{prop}(x,y,\bar{s})\},
\label{723.1}
\end{equation}%
where $v_{prop}$ is the function $v$ in (\ref{eq:v}) which is constructed
from the propagated measured data on the propagation plane $\Gamma _{p}$.
Note that the function $v_{prop}$ has a negative peak corresponding to each
target, see Figure \ref{fig:prep2}. The truncation value 0.85 was chosen
based on trial-and-error tests on some non-blind targets with known sizes.
We observed that $\Gamma _{T}$ provided a good approximation for the $xy$
projection of a target. Note that the same truncation was applied to blind
targets as well. Hence, it is not biased.

\begin{figure}[tph]
\centering
\subfloat[]{\includegraphics[width = 0.4\textwidth,height =
0.35\textwidth]{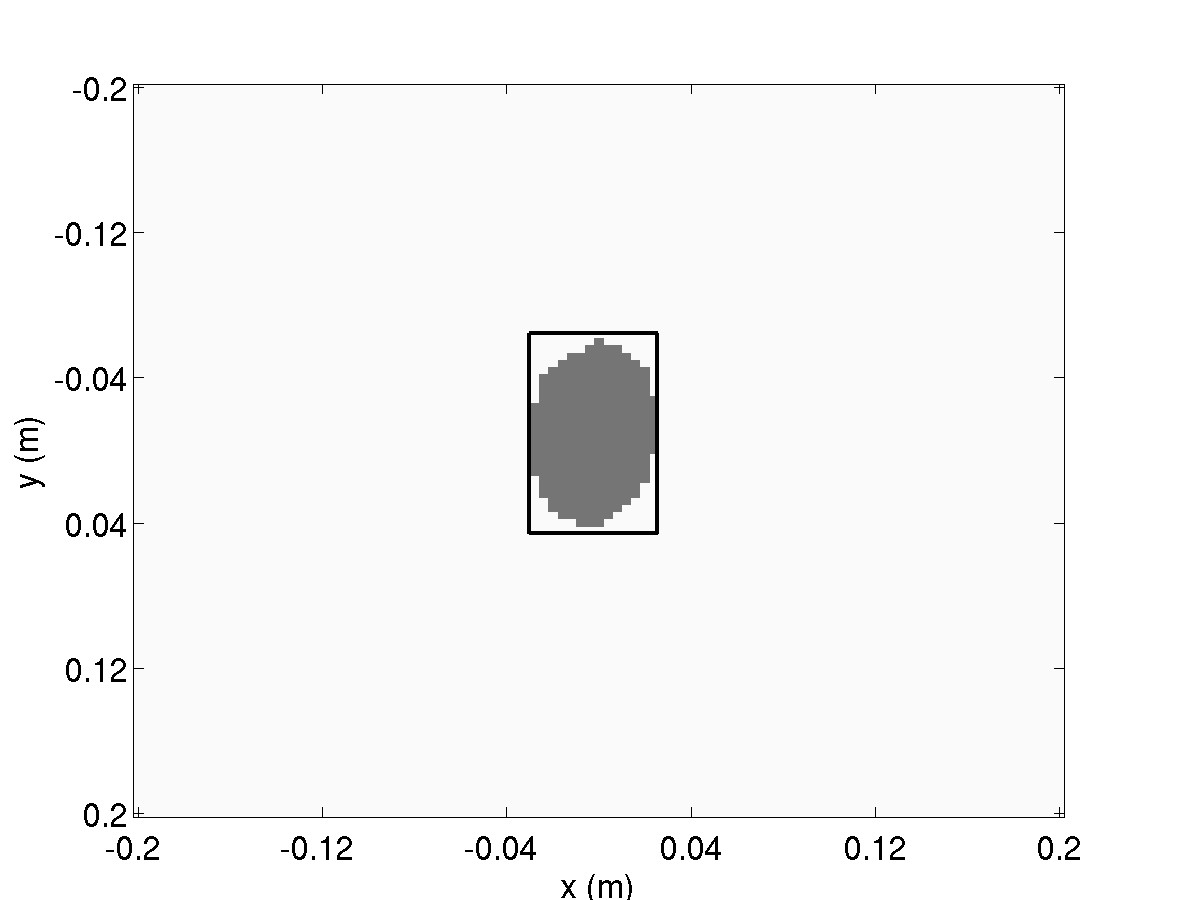}} 
\subfloat[]{\includegraphics[width = 0.4\textwidth,height =
0.35\textwidth]{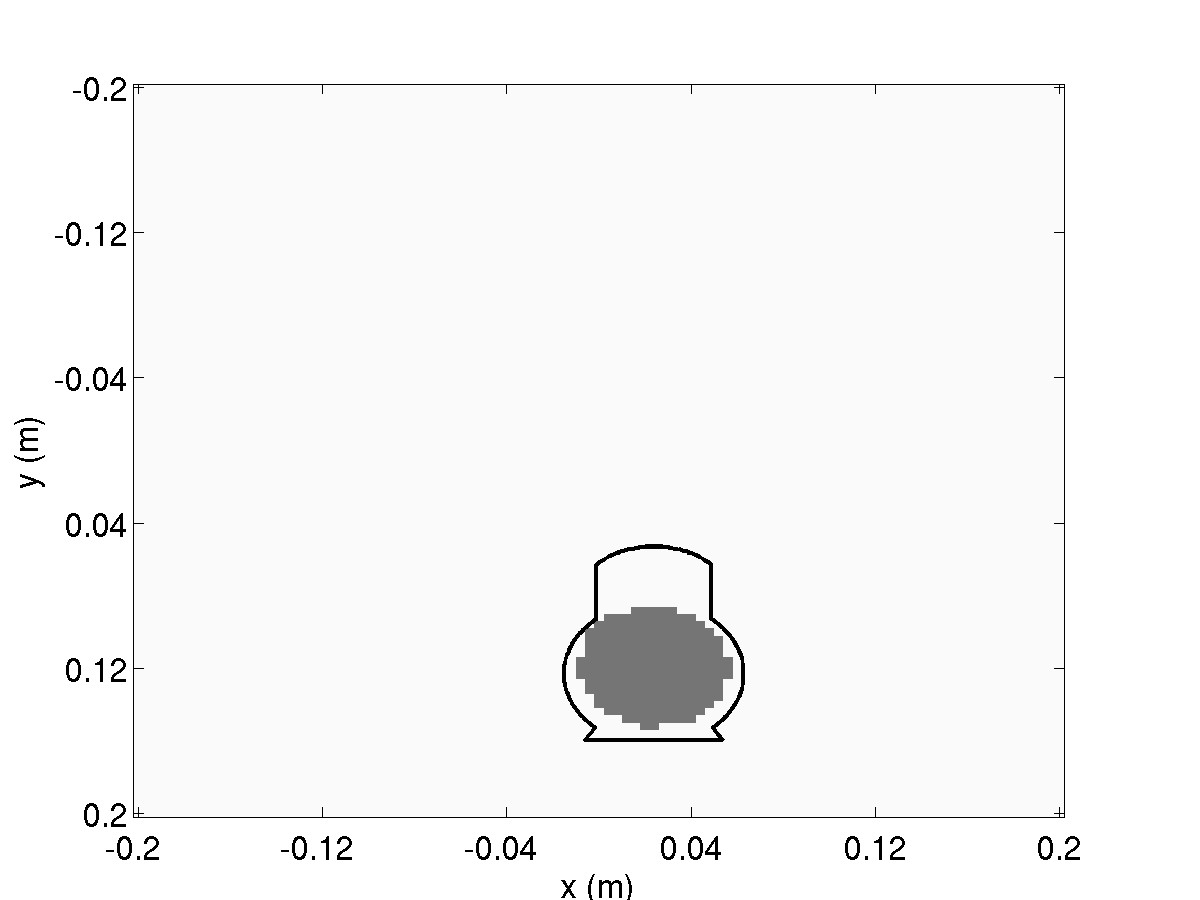}}
\caption{Estimation of target's $xy$ projection: (a): target 4 (a metallic
cylinder); (b): target 10 (a wooden doll partly filled with sand). Thin lines
indicate boundaries of correct $xy$ projections.}
\label{fig:size}
\end{figure}

Figure \ref{fig:size} shows the estimated $xy$ projections of targets 4 and
10 in our experiments, see Table \ref{table1}. Although the corners of the
targets are not well estimated, we see that their shapes and sizes are
reasonably good. For target 10, since its lower part was filled with the
sand and the upper part was air inside of a wooden cover (see section \ref%
{sec:8.1} for details), we can see only the lower part due to its higher
refractive index compared to the upper part.

\subsection{Complementing backscattered data}

We recall that only backscattered signals were measured in our experiments.
This means that after data propagation, the function $\psi (\mathbf{x},s)$
was known only on the side $\Gamma _{p}=\{\mathbf{x}\in \partial \Omega
:\left\vert x\right\vert ,z=0.04\}$ of $\Omega .$ As in \cite{B-K:JIIP2012},
we replace the missing data on the other parts of $\partial \Omega $ by the
corresponding solution of the forward problem in the homogeneous medium. In
other words, in our computations, function $\psi $ is given by 
\begin{equation}
\psi (\mathbf{x},s)=%
\begin{cases}
\psi _{prop}(\mathbf{x},s),\mathbf{x}\in \Gamma _{p},s\in (\underbar s,\bar{s%
}), \\ 
\psi _{sim}^{i}(\mathbf{x},s),\mathbf{x}\in \partial \Omega \setminus \Gamma
_{p},s\in (\underbar s,\bar{s}), \\ 
\end{cases}
\label{eq:funpsibs}
\end{equation}%
where $\psi _{prop}$ is the function $\psi $ of (\ref{eq:funpsi}) computed
from the propagated measured data at $\Gamma _{p}$ and $\psi _{sim}^{i}$ is
computed from the simulated solution of the problem (\ref{eq:fp01})--(\ref%
{eq:fp07}) with $\epsilon (\mathbf{x})\equiv 1$ in the wave equation (\ref%
{eq:fp01}). In the following, we also denote by $V_{prop}(\mathbf{x}%
):=v_{prop}(x,y,\bar{s})$, $\mathbf{x}\in \Gamma _{p},s\in \lbrack \underbar %
s,\bar{s}]$ the tail function (\ref{eq:tail}) of the propagated measured
data.

Recall that our measured data were collected with the step size of $0.02$ m
in $x$ and $y$ directions. To obtain the data at the same grid size as in
our computational domain, we applied the linear interpolation to the Laplace
transform of the propagated measured data.

Below we present the reconstruction results of two different tests: \textbf{%
Test 1} and \textbf{Test 2}. In Test 1, we made use of the first tail
function computed from the boundary value problem (\ref{3.37})--(\ref{3.38}%
). As it was remarked in section \ref{sec:gc}, the global convergence with
this choice of the initial tail function is rigorously guaranteed. In Test
2, we choose the first tail function from some information about the targets
which we obtained in the data pre-processing. Although the convergence of
the resulting algorithm has not been rigorously proved yet, our numerical
results show good reconstructions. As we mentioned in section \ref{subsec:alg},
stopping criteria of the algorithm should be addressed numerically. We will
discuss this in the following.

\subsection{Test 1}

When testing the algorithm for different non-blind targets in this test, we
have developed a reliable stopping criterion which includes two steps.

\textbf{Stopping criterion for inner iterations with respect to }$i$. The
inner iterations are stopped at $i=m_{n}$ such that either (\ref{7.5}) or (%
\ref{7.5_1}) is fulfilled, 
\begin{eqnarray}
&&E_{n,i}\geq E{_{n,i-1}}\text{ or }E{_{n,i}}\leq \eta ,  \label{7.5} \\
\text{ } &&\text{{}}D{_{n,i}}\geq D{_{n,i-1},}\text{ or }D{_{n,i}}\leq \eta ,
\label{7.5_1}
\end{eqnarray}%
where $\eta =10^{-6}$ is a chosen tolerance and $D_{n,i}=||{V}%
_{n,i}|_{\Gamma _{p}}-{V}_{prop}||_{L_{2}(\Gamma _{p})}.$ In (\ref{7.5}) $%
E_{n,i}$ represents the relative error of the coefficient, which is given by 
\begin{equation}
E_{n,i}=\frac{||{\epsilon }_{n,i}-{\epsilon }_{n,i-1}||_{L_{2}(\Omega )}}{||{%
\epsilon }_{n,i-1}||_{L_{2}(\Omega )}}.  \notag
\end{equation}

\begin{figure}[tph]
\centering
\subfloat[]{\includegraphics[width=0.48\textwidth,height =
0.38\textwidth]{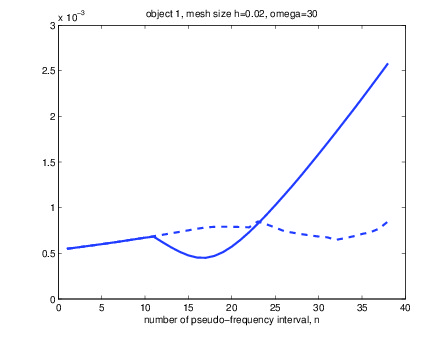}} 
\subfloat[]{\includegraphics[width=0.48\textwidth,height =
0.38\textwidth]{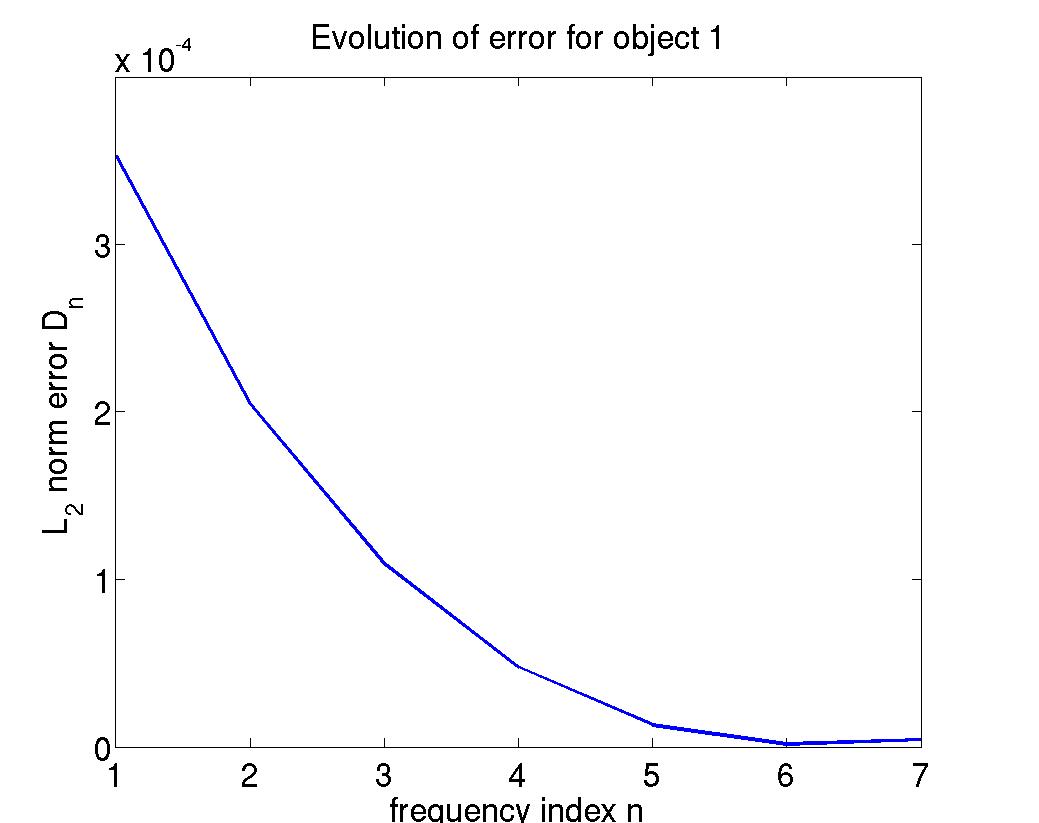}}
\caption{(a) Behavior of the norms $D_{n,first}$ (solid curve) and $%
D_{n,final}$ (dash curve) of Test 1; (b) Behavior of $D_{n,final}$ of Test 2. }
\label{fig:result0}
\end{figure}

\textbf{Criterion for choosing the final coefficient}. In Test 1, the
inversion algorithm was run for all pseudo frequencies in the chosen
interval $s\in \lbrack 8,10]$ and then the final reconstructed coefficient
was chosen as follows. On every pseudo frequency interval $\left[
s_{n},s_{n-1}\right) $ we took the norms $D_{n,first}=D_{n,1}$ and $%
D_{n,final}:=D_{n,m_{n}}$ at the first and the final inner iteration
respectively. We have always observed in all ten sets of our data that the
first norm $D_{n,first}$ increases first with respect to $n$, then decreases
and attains an unique minimum with respect to $n\in \left[ 1,N\right] =\left[
1,40\right] .$ On the other hand, the final norm $D_{n,final}$ has either
one or two local minima, see Figure \ref{fig:result0}(a). Let $n_{1}$ be the
number of the iteration $n$ at which the minimum of the first norms is
achieved, e.g. $n_{1}=16$ in Figure \ref{fig:result0}(a). As the
reconstructed coefficient $\epsilon _{rec}\left( \mathbf{x}\right) $, we
first choose $\epsilon _{rec}\left( \mathbf{x}\right) :=\epsilon
_{n_{1}}\left( \mathbf{x}\right) .$ Next, if $\max_{\overline{\Omega }%
}\epsilon _{rec}\left( \mathbf{x}\right) <5$ or $\max_{\overline{\Omega }%
}\epsilon _{rec}\left( \mathbf{x}\right) >10$, then we take the function $%
\epsilon _{rec}\left( \mathbf{x}\right) $ as the final reconstruction.
Suppose now that $5\leq \max_{\overline{\Omega }}\epsilon _{rec}\left( 
\mathbf{x}\right) \leq 10.$ Then we consider the iteration number $n_{2}$ at
which the smallest local minimum of the final norm $D_{n,final}$ is
achieved, e.g. $n_{2}=33$ in Figure \ref{fig:result0}(a). Then we take the
function $\epsilon _{rec}\left( \mathbf{x}\right) :=\epsilon _{n_{2}}\left( 
\mathbf{x}\right) $ as the final reconstruction.

\begin{figure}[tph]
\centering
\subfloat[Target 4, 3D
view]{\includegraphics[width=0.32\textwidth]{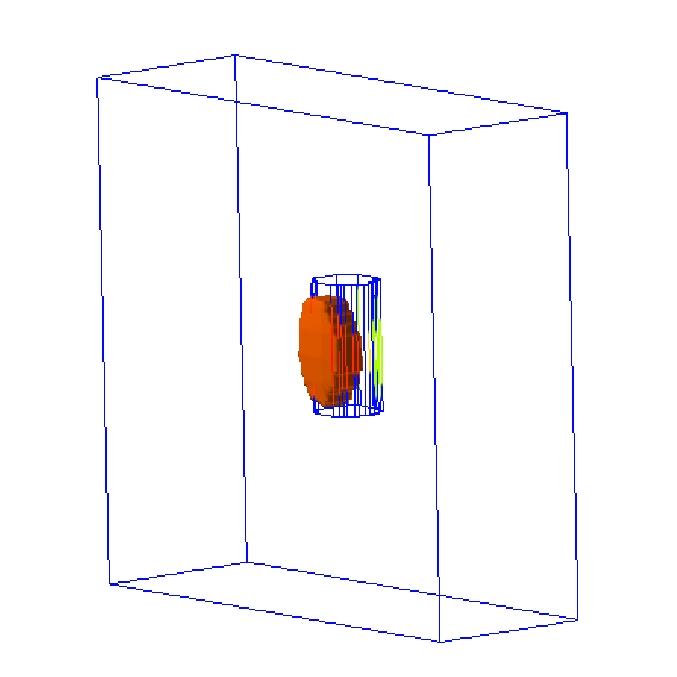}} 
\subfloat[Target 4, $xy$
view]{\includegraphics[width=0.32\textwidth]{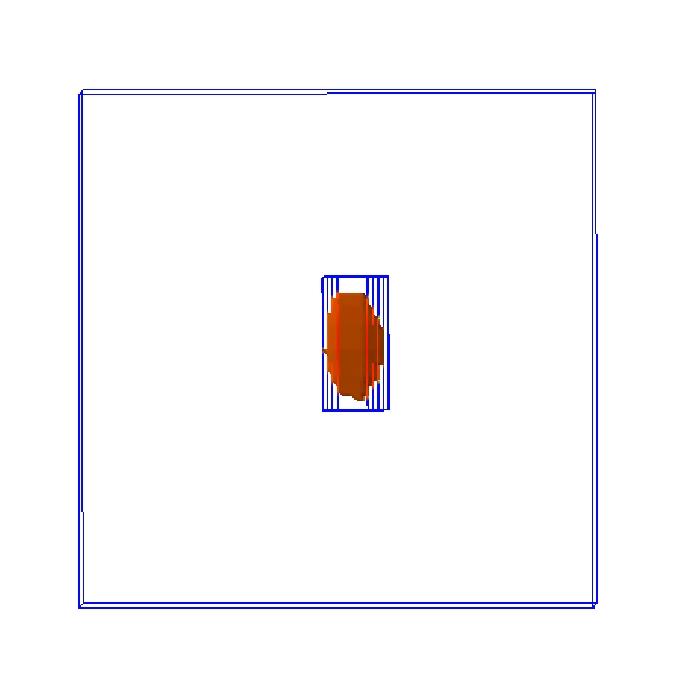}} 
\subfloat[Target 4, $yz$
view]{\includegraphics[width=0.32\textwidth]{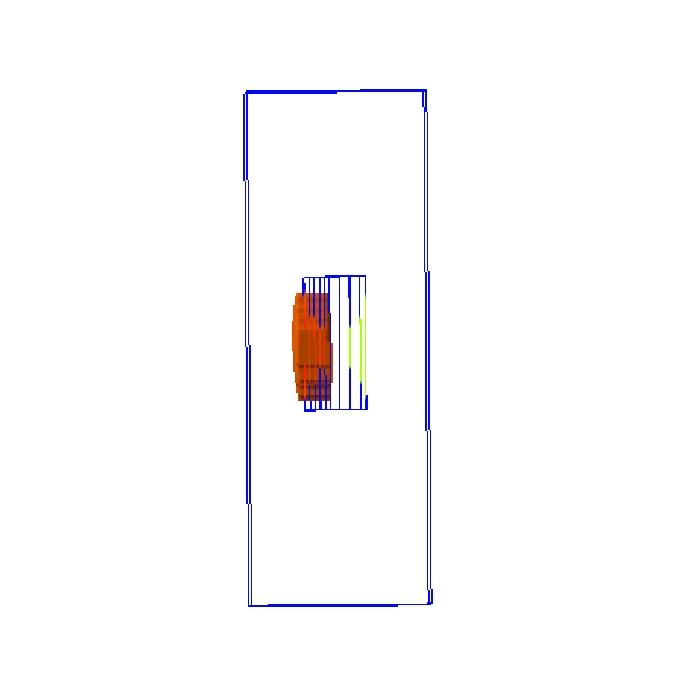}}
\par
\subfloat[Target 10, 3D
view]{\includegraphics[width=0.32\textwidth]{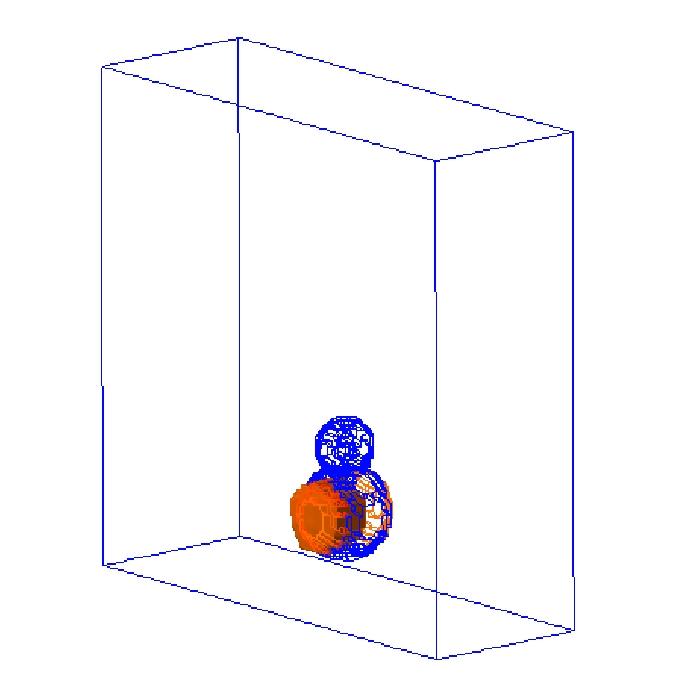}} 
\subfloat[Target 10, $xy$
view]{\includegraphics[width=0.32\textwidth]{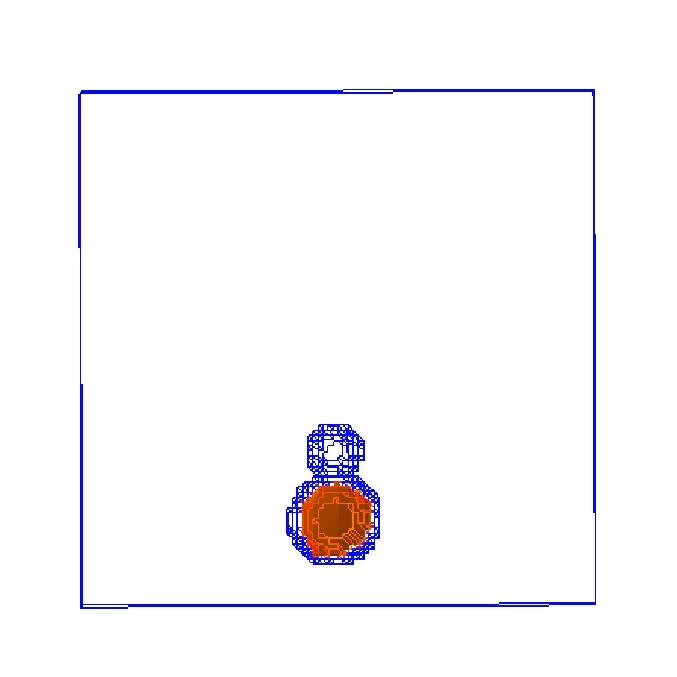}} 
\subfloat[Target 10, $yz$
view]{\includegraphics[width=0.32\textwidth]{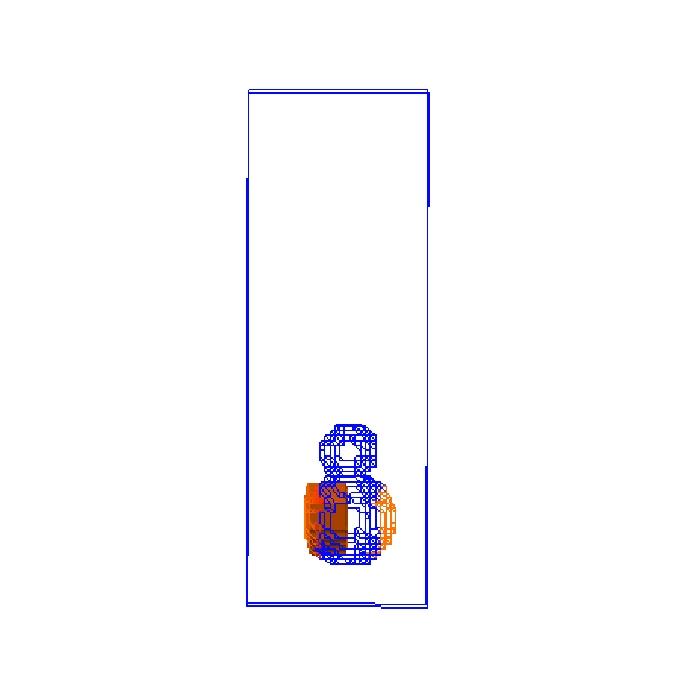}}
\caption{Some reconstruction results of Test 1. $xy$ view means the
projection of the target on the $xy$ plane. $yz$ view means the projection
of the target on the $yz$ plane. Thin lines indicate correct shapes.}
\label{fig:result1}
\end{figure}

As shown in Table \ref{table1}, the reconstructed refractive indices are
quite close to the true values for all dielectric targets. Table \ref{table2}
shows that reconstructed appearing dielectric constants of metallic targets
are always in the required range (\ref{X}). However, the shapes and sizes of
the targets were not well reconstructed, in particular the \textquotedblleft
depth" in the $z-$direction. To improve this, we use the following
post-processing procedure. Let $P_{z_{0}}:=\left\{ z=z_{0}\right\} $ be the
plane where the function $\epsilon _{rec}\left( \mathbf{x}\right) $ achieves
its maximal value. 
Then we compute the truncated coefficient function $\tilde{\epsilon}%
_{rec}\left( \mathbf{x}\right) $ as 
\begin{equation}
\tilde{\epsilon}_{rec}\left( \mathbf{x}\right) =%
\begin{cases}
\epsilon _{rec}\left( \mathbf{x}\right) \text{ if }\epsilon _{rec}\left(
x,y,z_{0}\right) >\gamma \max \epsilon _{rec}\left( x,y,z_{0}\right) , \\ 
1\text{ otherwise},%
\end{cases}
\label{A}
\end{equation}%
where $\gamma $ is a truncation factor chosen such that the area of $\{%
\widetilde{\epsilon }\left( x,y,z_{0}\right) >1\}$ is the same as that 
of $\Gamma _{T}$, see (\ref{723.1}) for $\Gamma _{T}$. Finally, we
approximate the depth in the $z$ direction by truncating $\tilde{\epsilon}%
_{rec}\left( \mathbf{x}\right) $ by 90\% of its maximal value. This
truncation value is chosen based on the trial-and-error tests with non-blind
targets. Figure \ref{fig:result1} shows imaging results for targets 4 and 10.

\subsection{Test 2}

In this test, we chose the mesh size of $0.01$ by $0.01$ by $0.01$ in order
to represent the targets' shapes more accurately. We use an information
about the targets in our data pre-processing to restrict the estimated
coefficient $\epsilon $ in a subdomain of $\Omega $ and to choose the first
tail function. More precisely, for each target, let $x_{t,min}=\min \{x\in
\Gamma _{T}\},\quad x_{t,max}=\max \{x\in \Gamma _{T}\}.$ The numbers $%
y_{t,min}$ and $y_{t,max}$ are defined similarly. Then, we define the
extended projection by 
\begin{equation*}
\Gamma _{T,ext}=\{x_{t,min}-0.03\leq x\leq x_{t,max}+0.03,\
y_{t,min}-0.03\leq y\leq y_{t,max}+0.03\}.
\end{equation*}%
Moreover, denote by $z_{t,front}$ the estimated location of the front side
of the target in the $z$ direction. We then define the following domain $%
\Omega _{T,ext}$ 
\begin{equation*}
\Omega _{T,ext}:=\{\mathbf{x}\in \Omega :(x,y)\in \Gamma _{T,ext},\ -0.1\leq
z\leq z_{t,front}\}.
\end{equation*}%
Clearly, $\Omega _{T,ext}\subset \Omega $. Moreover, this domain should contain the unknown target we are looking
for. Next, we chose the first tail function $V_{0}$ as the function (\ref%
{eq:tail}), where the function $w\left( \mathbf{x},\overline{s}\right) $ is
the Laplace transform (\ref{eq:laplacetr}) at $s=\overline{s}$ of the
solution of the forward problem (\ref{eq:fp01})--(\ref{eq:fp07}) with the
coefficient $\epsilon =\epsilon _{0}$, where 
\begin{equation*}
\epsilon _{0}(\mathbf{x})=1+d,\text{ for }\mathbf{x}\in \Omega _{T,ext},\
\epsilon _{0}(\mathbf{x})=1,\text{ for }\mathbf{x}\notin \Omega _{T,ext}.
\end{equation*}%
Moreover, the coefficient is truncated by
\begin{equation}
\epsilon _{n,i}(\mathbf{x})=%
\begin{cases}
\epsilon _{n,i}(\mathbf{x}) & \text{ if }\mathbf{x}\in \Omega _{T,ext}\text{
and }1\leq \epsilon _{n,i}(\mathbf{x})\leq 1+d, \\ 
1, & \text{ if } \epsilon _{n,i}(\mathbf{x}) <  1,\\
1 + d,  & \text{ if } \epsilon _{n,i}(\mathbf{x}) > 1 + d.
\end{cases}
\label{eq:coeftrunc2}
\end{equation}%
In this paper, we chose $1+d$ to be 10 for nonmetallic targets and 20 for
metallic ones.

\textbf{Stopping criterion.} In this test, the inner iterations were stopped
using the same criterion as in Test 1. However, we also set the maximum
number of inner iterations to be 5. That means, the inner iterations were
stopped if either (\ref{7.5}) or (\ref{7.5_1}) was satisfied for $i<5,$ or
if $i=5$.

Concerning the outer iterations, we have observed that the error $%
D_{n,final}$ decreased with respect to $n$ first, and then increased after
reaching a minimum, see Figure \ref{fig:result0}(b). At that minimum, the
estimated coefficient was a good approximation of the true one for some
non-blind targets. Therefore, we stopped the algorithm when this error
function attained the minimum.

\begin{figure}[tph]
\centering
\subfloat[Target 4, 3D
view]{\includegraphics[width=0.32\textwidth]{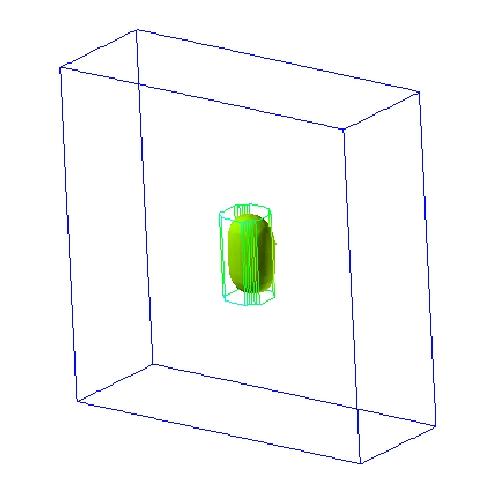}} 
\subfloat[Target 4, $xy$
view]{\includegraphics[width=0.32\textwidth]{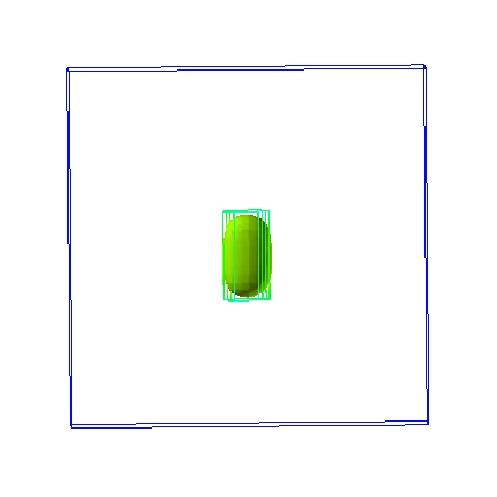}} 
\subfloat[Target 4, $yz$
view]{\includegraphics[width=0.32\textwidth]{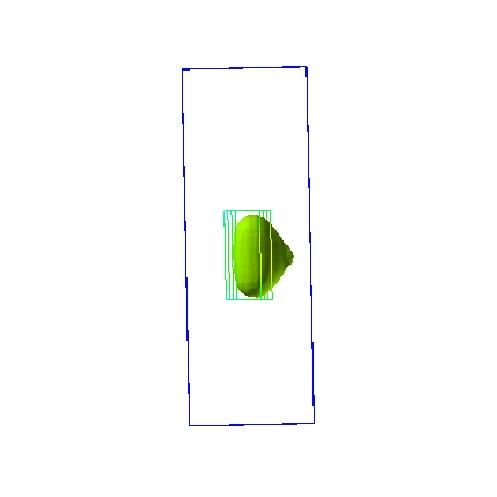}}
\par
\subfloat[Target 10, 3D
view]{\includegraphics[width=0.32\textwidth]{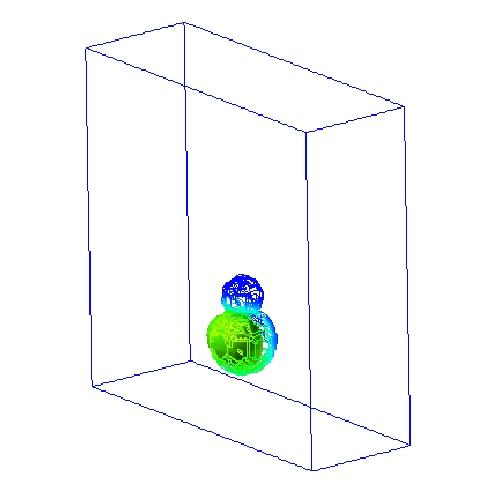}} 
\subfloat[Target 10, $xy$
view]{\includegraphics[width=0.32\textwidth]{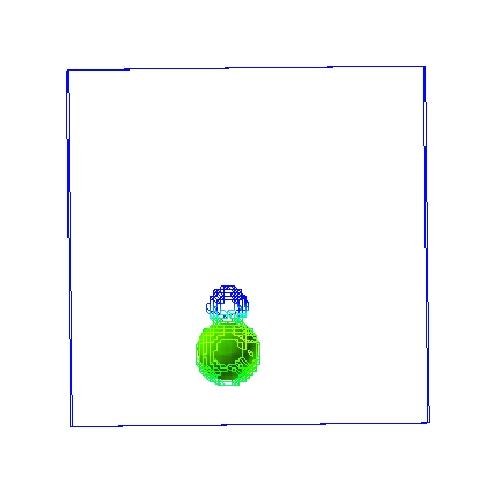}} 
\subfloat[Target 10, $yz$
view]{\includegraphics[width=0.32\textwidth]{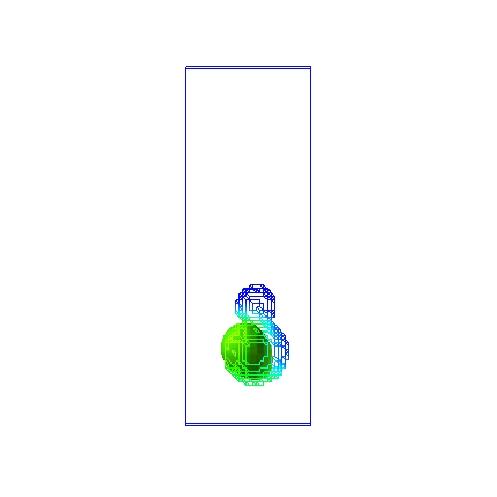}}
\caption{Some reconstruction results of Test 2. Thin lines indicate correct
shapes.}
\label{fig:result2}
\end{figure}

We have observed through our tests that the shapes of the targets are quite
well reconstructed. Figure \ref{fig:result2} shows the results of targets 4
and 10 using Test 2. For target 10, we again obtained the lower part, which
was filled with the sand, see Figure \ref{fig:size}(c) for its $xy$
projection estimated from the data (recall that air was inside the wooden
cover in the upper part of that target).

\begin{table}[tph]
\caption{{\protect\small \emph{Computed and measured refractive indices of
six dielectric targets. The average error in direct measurements is 24\%.
The average error of Test 1 is 8\% and Test 2 is 3.4\%.}}}
\label{table1}
\begin{center}
{\footnotesize 
\begin{tabular}{|l|l|l|l|l|l|}
\hline
Target ID & 1 & 2 & 5 & 8 & 10 \\ \hline
blind/non-blind (yes/no) & no & no & yes & yes & yes \\ \hline
Measured value, error & 2.11, 19\% & 1.84, 18\% & 2.14, 28\% & 1.89, 30\% & 
2.1, 26\% \\ \hline
$n^{comp}$ Test 1, error: & 1.92, 10\% & 1.80, 2\% & 1.83, 17\% & 1.86, 2\%
& 1.92, 9\% \\ \hline
$n^{comp}$ Test 2, error: & 2.03, 4\% & 1.96, 7\% & 2.10, 2\% & 1.85, 2\% & 
2.05, 2\% \\ \hline
\end{tabular}
}
\end{center}
\end{table}

\begin{table}[tph]
\caption{{\protect\small \emph{Computed appearing dielectric constants }$%
\protect\epsilon ^{comp}$\emph{\ of metallic targets.}}}
\label{table2}
\begin{center}
{\footnotesize 
\begin{tabular}{|l|l|l|l|l|l|}
\hline
Target ID & 3 & 4 & 6 & 7 & 9 \\ \hline
blind/non-blind (yes/no) & no & no & yes & yes & yes \\ \hline
$\epsilon^{comp}$ Test 1: & 14.37 & 16.93 & 25.0 & 13.61 & 13.56 \\ \hline
$\epsilon^{comp}$ Test 2: & 7.59 & 10.76 & 19.55 & 8.12 & 7.89 \\ \hline
\end{tabular}
}
\end{center}
\end{table}

\subsection{Summary of reconstruction results}

\label{sec:8.1}

To compare our computational results with true ones, we have directly
measured \emph{a posteriori} refractive indices $n=\sqrt{\epsilon }$ of
dielectric targets. In Tables \ref{table1}, \ref{table2}, the computed
values of the refractive index $n^{\text{comp}}$ of dielectrics
(respectively appearing dielectric constant $\epsilon ^{\text{comp}}$ of
metals) were chosen as the square root of the maximal values (respectively,
the maximal values) of the reconstructed coefficient. Table \ref{table1}
lists both computed $n^{\text{comp}}$ and directly measured refractive
indices $n$ of dielectric targets for both Test 1 and Test 2. This table
also shows the measurement error in direct measurements of $n$. Table \ref%
{table2} lists computed appearing dielectric constants $\epsilon ^{\text{comp%
}}$ of metallic targets. We see from Tables \ref{table1} and \ref{table2}
that $\left( n^{\text{comp}}\right) ^{2}<5$ for all dielectric targets,
while $\epsilon ^{\text{comp}}>13.56$ for all metallic targets in Test 1 and $%
\epsilon ^{\text{comp}} \ge7.59$ in Test 2. Thus, our algorithm can
differentiate quite well between dielectric and metallic targets.

It can be seen from Table \ref{table1} that both tests image refractive
indices of both blind and non-blind dielectric targets with only a few
percent of error, which is even smaller than the error of direct
measurements. The average error of computed refractive indices $n^{\text{comp%
}}$ in Tests 1 and 2 is respectively three and seven times less than the
average error of direct measurements. Test 1 obtains higher appearing
dielectric constant of some metallic targets than Test 2. However, Test 2
provides better shapes.

Unlike targets 1 - 7, which are homogeneous, targets 8, 9 and 10 are
heterogeneous. Target 8 is a wooden doll with air inside. Target 9 is that
doll with a piece of metal inside, i.e. this is a metal covered by a
dielectric. We can see that only the metal was imaged, because its
reflection is much stronger than that of the wood. Target 10 is the same
doll partially filled with sand inside (except of the top), i.e.~this is one
dielectric covered by another one. One can see that only the part with the
sand was imaged in target number 10, since its dielectric constant is higher
than the air inside the top. Moreover, the reconstructed refractive index is
about the average of those of the wood and the sand.

In conclusion, we can see from our tests that, with the proposed data
pre-processing procedure, the globally convergent algorithm can image quite
well both geometries and materials of the targets in our experimental data
even though there is a huge misfit between these data and simulations.
Moreover, it can image large inclusion/background contrasts, the case that
is well known to be difficult for conventional least squares approaches.

\section*{Acknowledgment}
This research was supported by US Army Research Laboratory and US Army
Research Office grants W911NF-11-1-0325 and W911NF-11-1-0399, the Swedish
Research Council, the Swedish Foundation for Strategic Research (SSF) in
Gothenburg Mathematical Modelling Centre (GMMC) and by the Swedish
Institute, Visby Program.
The authors are grateful to Mr. Steven Kitchin for his excellent work on
data collection. %


\end{document}